\pdfoutput=1
\documentclass[11pt,a4paper]{article}
\usepackage{jinstpub}
\usepackage[utf8]{inputenc}
\usepackage[T1]{fontenc}
\usepackage[english]{babel}
\usepackage{lineno}
\usepackage{amsmath,amssymb}
\usepackage{graphicx}
\usepackage{xspace}
\usepackage{units}
\usepackage{booktabs}
\usepackage[capitalize]{cleveref}
\usepackage{xcolor}
\usepackage{enumitem}

\graphicspath{{figs/}}

\def\Offline{\mbox{$\overline{\textrm{Off}}$\hspace{.05em}\protect\raisebox{.4ex}{$\protect\underline{\textrm{line}}$}}\xspace}

\title{Studies on the response of a water-Cherenkov detector of the Pierre Auger Observatory to atmospheric muons using an RPC hodoscope}

\emailAdd{auger\_spokespersons@fnal.gov}

\abstract{
Extensive air showers, originating from ultra-high energy cosmic rays, have been successfully measured through the use of arrays of water-Cherenkov detectors (WCDs).
Sophisticated analyses exploiting WCD data have made it possible to demonstrate that shower simulations, based on different hadronic-interaction models, cannot reproduce the observed number of muons at the ground.
The accurate knowledge of the WCD response to muons is paramount in establishing the exact level of this discrepancy.
In this work, we report on a study of the response of a WCD of the Pierre Auger Observatory to atmospheric muons performed with a hodoscope made of resistive plate chambers (RPCs), enabling us to select and reconstruct nearly 600 thousand single muon trajectories with zenith angles ranging from $0^\circ$ to $55^\circ$.
Comparison of distributions of key observables between the hodoscope data and the predictions of dedicated simulations allows us to demonstrate the accuracy of the latter at a level of 2\%. 
As the WCD calibration is based on its response to atmospheric muons, the hodoscope data are also exploited to show the long-term stability of the procedure.
}


\author{\includegraphics[height=30mm]{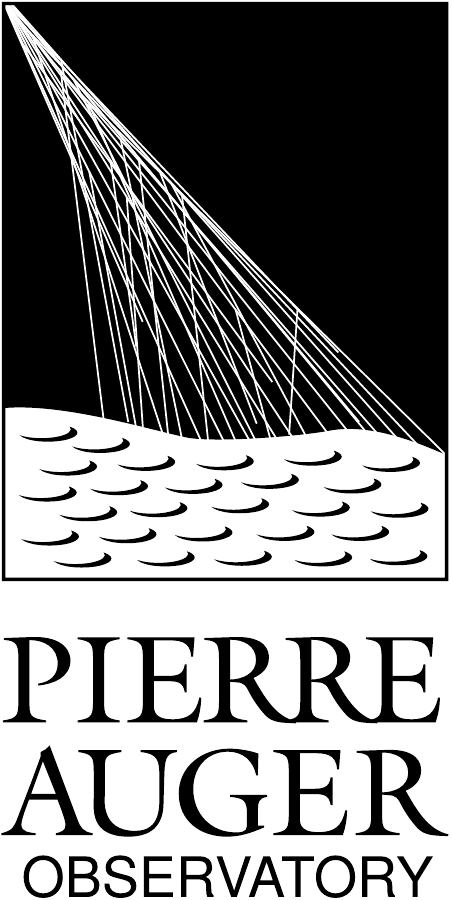}\\[3mm]The Pierre Auger Collaboration}
\affiliation{Av.\ San Mart\'{\i}n Norte 306, 5613 Malarg\"ue, Mendoza, Argentina}

\begin{document}

\maketitle

\section{Introduction}
\label{SecIntroduction}

The Pierre Auger Observatory, located at an altitude of ${\sim}1400$\,m above sea level near Malarg\"ue in the province of Mendoza, Argentina, is the largest facility in the world dedicated to the detection of ultra-high energy cosmic rays (UHECRs) in the energy range from ${\sim}10^{17}$ eV up to the highest energies~\cite{Auger-NIM-A}.
Due to the very low flux at these energies, the observation of UHECRs is performed indirectly by recording the extensive air showers produced by these particles when they interact in the atmosphere.

At the Pierre Auger Observatory, extensive air showers are observed using two detection techniques.
Telescopes collecting the fluorescence light emitted by atmospheric nitrogen, excited after the passage of the charged particles, allow for the observation of the longitudinal profile of the showers.
This technique provides a nearly calorimetric estimate of the energy carried by the primary particle.
However, this technique is constrained to nights with low background light conditions, limiting its uptime to below 15\%.
The second detection technique uses a surface detector (SD) array composed of 1660 water-Cherenkov detectors (WCDs) deployed on the ground, in which the light produced in the water by charged particles above the threshold for emitting Cherenkov radiation is collected by three photomultiplier tubes (PMTs).
The SD operates with a duty cycle close to 100\%.
The detected signals in the SD are used to determine the arrival direction and to estimate the size of the showers.
The shower size of all such events is subsequently converted into the energy of the primary cosmic ray through a calibration based on a subset of events detected by both the surface and fluorescence detectors.
This ``hybrid'' approach allows for a calorimetric estimate of the energy also for events recorded during periods when the fluorescence detector cannot be operated.

The detection and reconstruction of air showers allows not only for studies of the astrophysics of UHECRs, but also represents a unique opportunity to access particle interactions at energies that are far higher than could be achieved by any Earth-based accelerator.
The number of muons in showers is particularly sensitive to hadronic interactions taking place during the development of the cascade in the atmosphere.
Over the last 20 years, increasing numbers of studies (see~\cite{Cazon2019} for a recent review), including the Pierre Auger Observatory, have provided data showing indications of a discrepancy between the number of muons predicted in showers by different hadronic-interaction models and that observed in data.
In Auger Observatory, the analyses developed in this context are based on the data from WCDs, from which a muon deficit has been revealed in simulations at energies around and above $10^{19}$\,eV~\cite{HASPRD2015,GlennysPRL2016}. 

In the comparison between the observed showers and showers predicted by models, the detailed simulation of the WCD, which includes all the relevant physics processes, accounts for the detector geometry, and simulates the response of the electronics, naturally plays a crucial role.
The objective of this work is to probe experimentally this simulation in terms of the response to atmospheric particles, most notably background muons, at different zenith angles.
For this purpose, we have designed and deployed a hodoscope composed of resistive-plate chambers (RPCs), which, installed on one of the WCDs, enables the selection of single muons passing through the detector.
The RPC segmentation allows us to reconstruct muon trajectories and impact points, thus enabling the study of the signal response of the WCD for different zenith angles (from $0^\circ$ up to $55^\circ$) of arriving muons and the comparison with signals predicted by the detector simulation.
In addition, the operation of the hodoscope allows us to verify a component of the WCD calibration procedure~\cite{NIM-A-Calib}, which relies on the determination of the charge deposited by a vertical and centrally through-going atmospheric muon.
As the WCD is not a directional detector, the peak in the charge distribution for vertical centered-muons is obtained by scaling the peak in the charge distribution obtained with the omni-directional muons.
The latter is evaluated every minute for all data-taking WCDs, while the scaling factor was measured by means of a dedicated muon telescope on a reference WCD at the beginning of the operation of the Observatory~\cite{ProcCalib2005,NIM-A-Calib}. 
We took advantage of the RPC hodoscope to repeat with higher precision such a measurement and validate the scaling factor.

Overall, two data acquisition campaigns took place with the RPC hodoscope: one to detect muons with more inclined zenith angles (up to $55^\circ$) and the other dedicated to near-vertical muons.
The presentation of the measurements, of the data analysis, and of the results is organised as follows.
In \cref{SecExperimentalSetup}, we first describe the experimental setup, including a brief description of the features of the WCD, the RPC specifications, the related electronics and trigger system, as well as the different acquisition configurations adopted and the data obtained.
The following \cref{SecSimulation} illustrates the characteristics of the generated showers and the characteristics of the detector simulation.
As for the latter, in \cref{SecSimulationParameters} we provide a list of the most relevant parameters and their values.
In \cref{SecDataAnalysis}, we explain how the hodoscope data are used to select specific muon geometries and how the associated charge and trajectory are reconstructed.
Then we show that the distributions of these basic observables are comparable with those of the simulations and proceed to study, in \cref{SecResultsResponse}, the detailed response of the WCD to muons, down to the level of single PMTs.
In \cref{SecResultsVEM}, we then present the result of the new measurement of the scaling factor of the calibration before concluding in \cref{SecConclusions}.

\section{Experimental setup}
\label{SecExperimentalSetup}

\subsection{RPC hodoscope}
\label{SecRPC}

We set up an RPC hodoscope around a WCD located in the central campus of the Pierre Auger Observatory in Malarg\"ue.
This is one of the reference WCDs used for tests and verifications, as well as for the determination of the scaling factor for the calibration of the SD signals.
Like all other WCDs, it is a plastic cylinder with a 10\,m$^2$ base surface filled with ultra-pure water to the depth of 1.2\,m and an inner reflective liner made of Tyvek$^\text{\textregistered}$.
Floating on the top of the water surface and pointing downwards, three 9-inch PMTs are placed in a formation of an equilateral triangle with each PMT at a distance of 1.2\,m from the center.
Each PMT has two outputs: the low-gain signal, taken directly from the anode, and the high-gain signal, provided by the last dynode and amplified to be nominally 32 times larger than the low gain.
The low- and high-gain signals are digitised using 10-bit Flash Analog to Digital Converters (FADCs) with a sampling frequency of 40\,MHz.
A detailed description of the WCD can be found in~\cite{Auger-NIM-A}.

The RPCs used in this work (one above and the other below the WCD) were built for the Pierre Auger Observatory with the goal of providing high-accuracy muon measurements~\cite{EPJ-C-MARTA}, which demanded the development of autonomous units, reliably operating outdoors with high efficiency, and low gas (tetrafluoroethane) and power consumption~\cite{LL_RPC}.

The sensitive volume of an RPC is established by two 1\,mm gas gaps formed between glass plates of 2\,mm thickness.
Outside this chamber, a high voltage is applied through a resistive layer.
The chamber itself is enclosed in an acrylic box that isolates the detector components (for instance, the gas and HV) from the exterior.

When the primary ionization and avalanche multiplication occur due to the passage of a charged particle through the sensitive volume, the signal from the electron avalanche (fast charge) is picked-up by induction on metallic plates placed on top of the gaseous volume.
These plates consist of an $8{\times}8$ matrix of pickup electrode pads, each with an area of $18{\times}14$\,cm$^2$.
The pads are separated by a guard ring which is connected to the ground potential and forms 1\,cm gaps between the pads.
The total area of the RPC is $1.2{\times}1.5$\,m$^2$.

The RPCs are operated in avalanche mode to minimize the occurrence of streamers, which can compromise the detector performance in the long term.
The signal from each pad is amplified and, to keep the electronics simple, a threshold discrimination is applied.
In this way, a digital signal is generated from the volume next to each pad where a particle has crossed the corresponding gas gap.
By analyzing active pads in an event, it is possible to locate the particle traversal with an uncertainty of approximately 7\,cm.

\begin{figure}[t]
\centering
\includegraphics[width=0.6\textwidth]{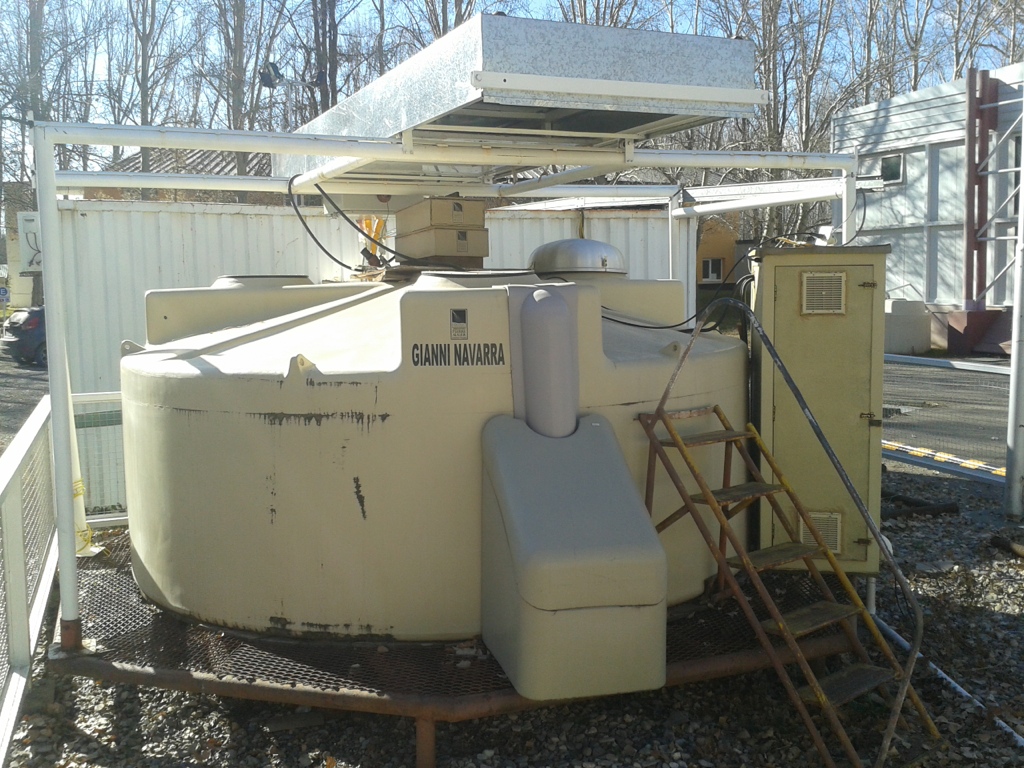}
\caption{A photograph of the experimental setup in Malarg\"ue, showing the aluminum box with the top RPC installed above the Gianni Navarra WCD.
The bottom RPC is under the mesh support structure of the WCD and is not directly visible.}
\label{fig:setup}
\end{figure}

A photograph of the experimental setup is shown in~\cref{fig:setup}.
The bottom RPC was installed below the mesh structure supporting the WCD, about 25\,cm below its base, and the top RPC was installed about 60\,cm above the top of the WCD with an orientation perpendicular to the bottom RPC.
By using rails and a cart, both RPCs could be easily displaced to change the hodoscope geometry and thereby select different zenith-angle ranges of crossing muons, with a resolution of $1^\circ$ in zenith angle (see \cref{fig:setup2}).

\subsection{DAQ and trigger}
\label{SecDAQTrigger}

The data acquisition system of the test WCD uses the standard Auger SD electronics~\cite{Auger-NIM-A} except that an external trigger is used and the control and readout are performed through a direct serial link to the console of the electronics microprocessor.
When a trigger is received from the RPCs, 19.2\,$\upmu$s (768 bins) of low- and high-gain FADC traces are stored in memory for offline processing. 
An internally generated trigger used by all data-taking WCDs, with a threshold corresponding to ${\sim}0.1$\,VEM (VEM stands for vertical-equivalent muon units), is used to collect data that are locally processed to produce calibration and monitoring histograms for the high-gain outputs. 
A total of 10 calibration histograms are thus acquired for each event, namely the ADC baselines (3), signal heights (3), signal charge (3), and signal charge for the sum of the PMTs (1). 

The RPC electronics are based on the prototype discrete-electronics system PREC (Prototype Readout Electronics -- Classic version), which uses an architecture with one motherboard and 12 front-end boards with 8 channels each~\cite{JINST-PREC}.
The signal from an RPC pad is amplified in a dedicated front-end channel and a simple, programmable threshold is applied to perform a 1-bit digitization.
The threshold is set just above the baseline to avoid the electronics noise.
The RPC event data thus consists of the 1-bit status of each pad, which gives an indication of whether it was hit or not.

The motherboard employs purely digital electronics composed of 14 field-programmable gate arrays (FPGAs) organized in a mother-daughters configuration.
The daughter FPGAs receive the digital signals from the front-end boards and reformat the signals to have a specified width in time.
A count of the number of hits is recorded internally to estimate the rate of each pad.
To indicate an activity in the group of pads, the daughter FPGA also outputs a logical-OR operation of the inputs for corresponding trigger generation.
The mother FPGA conveys the clock and trigger signal to all daughters.
To define a coincidence, it was possible to implement a trigger built from the daughter-FPGA information since there is a bus in the motherboard with 13 lines that connect to all FPGAs (mother and daughters).
One daughter FPGA was used to perform the trigger algorithm.
A trigger is generated when there is at least one active pad in both the top and the bottom RPC within a time window of $2{\times}500$\,ns.
The trigger is generated immediately after the condition is met.

Whenever the system triggers, two signals are generated:
(1) one trigger signal is sent to the daughter FPGAs directly, causing them to latch the state of the pads;
(2) another trigger signal is sent to the WCD, leading to the recording of the PMT traces and generating the timing and identifier of the event.

\subsection{Acquisition campaigns}
\label{SecCampaigns}

\begin{figure}[t]
\def\figh{0.46}
\centering
\includegraphics[height=\figh\textwidth]{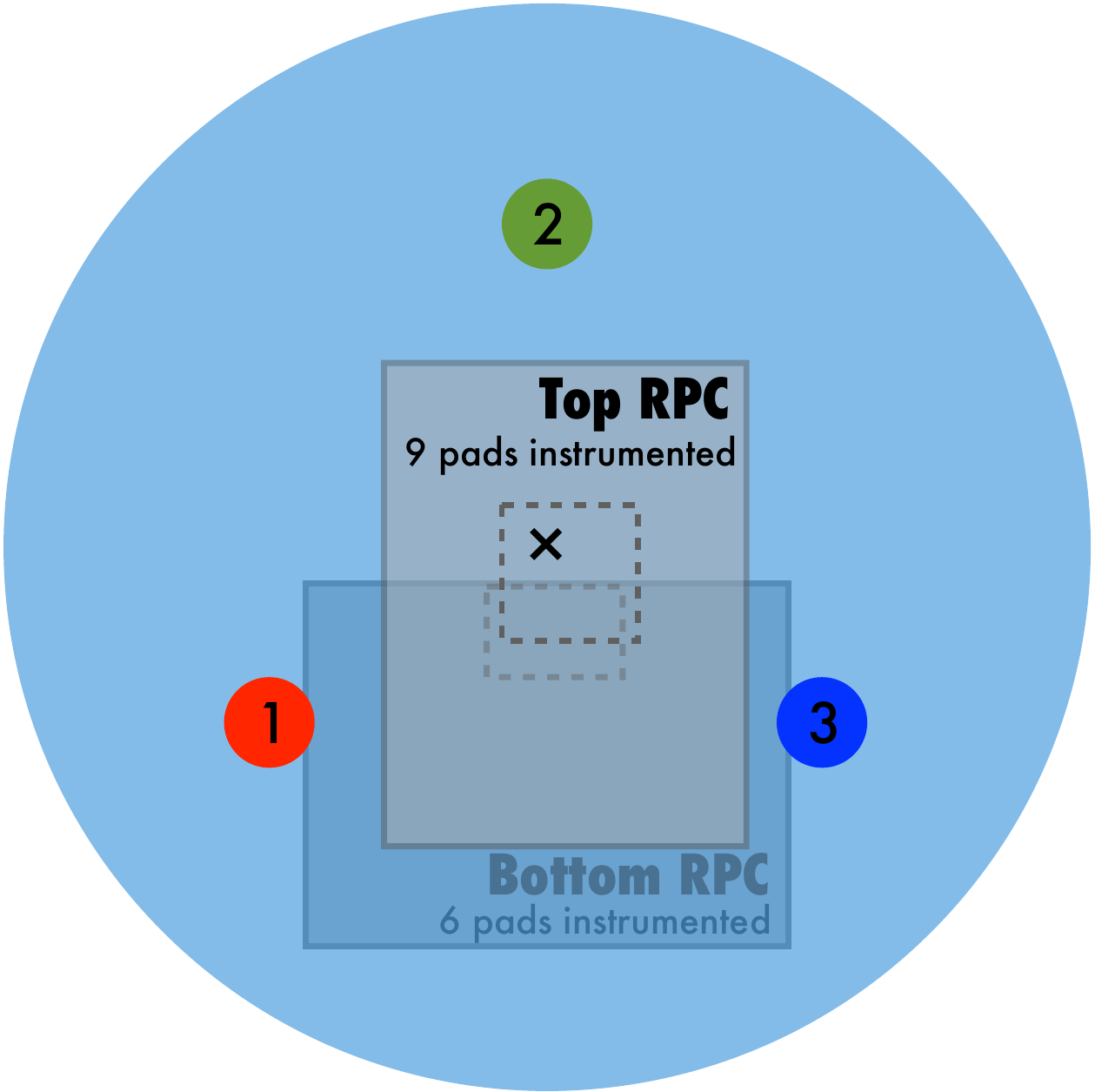}\hfill
\includegraphics[height=\figh\textwidth]{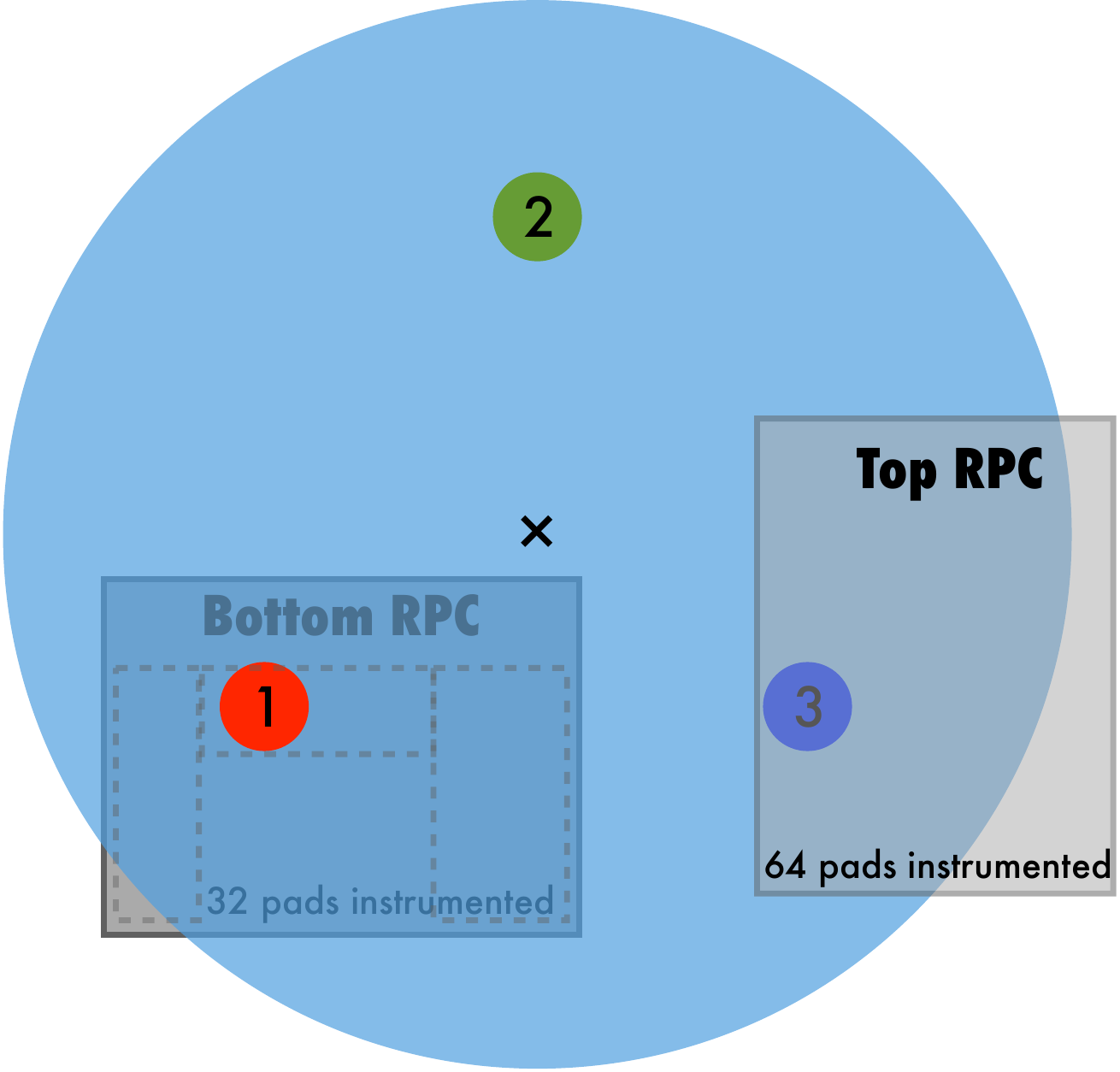}\hfill
\includegraphics[height=\figh\textwidth]{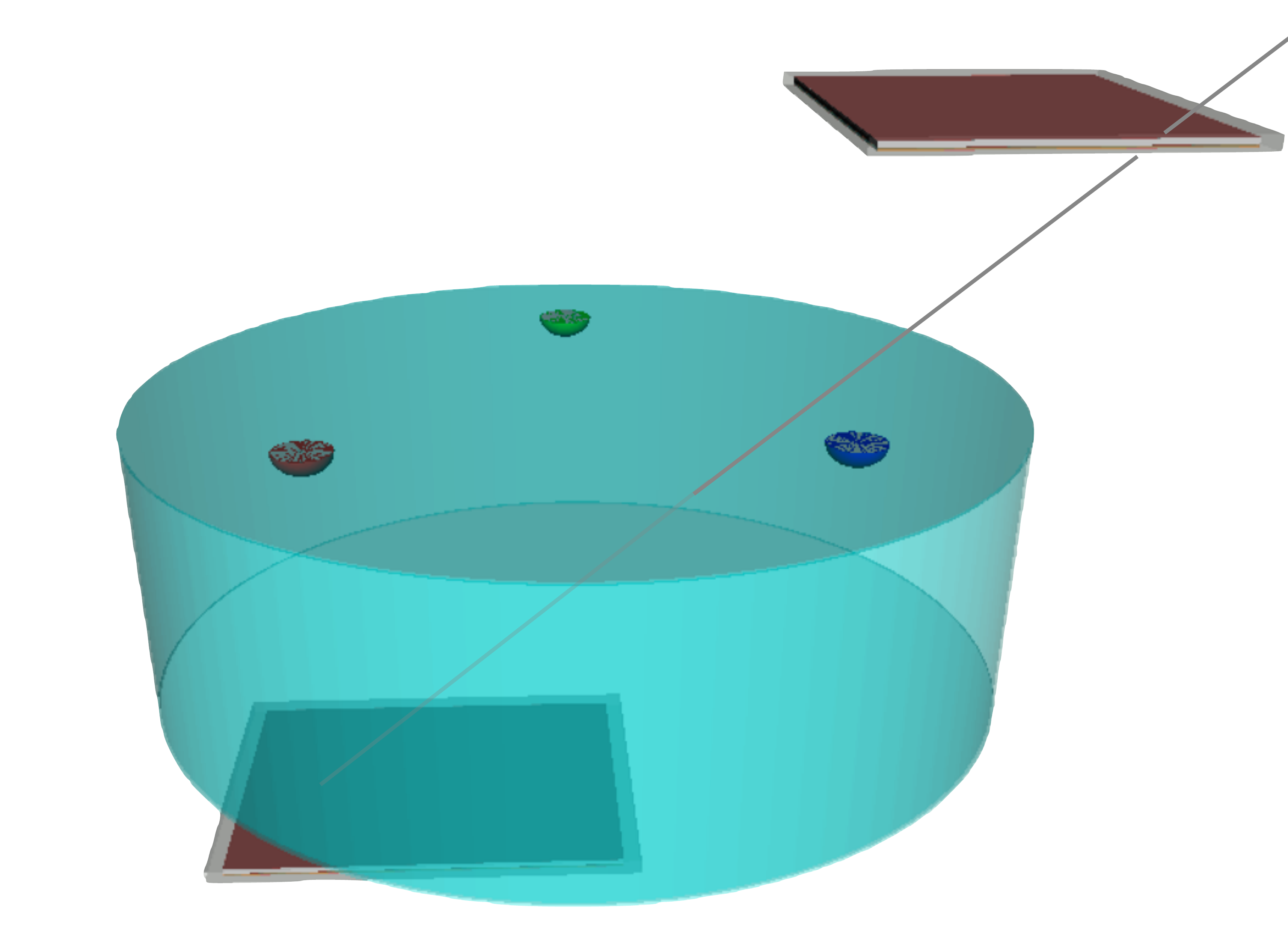}
\caption{A schematic overhead view of the experimental setup, with the hodoscope configuration for the acquisition of near-vertical muons (left) and for muons with zenith angles up to $55^\circ$ (right).
The dashed lines correspond to the narrowed active regions of RPCs.
The locations of the three PMTs are represented with the numbered circles.
The bottom figure shows a 3D scheme of the inclined setup and a muon trajectory.
While the less inclined muons enter the tank close to PMT\,3, the more inclined muons exit the tank close to PTM\,1.}
\label{fig:setup2}
\end{figure}

Two data acquisition campaigns have been performed:
(i) the measurement of muons with zenith angles up to ${\sim}55^\circ$, and
(ii) the measurement of near-vertical muons.
Both of them had an initial period for commissioning followed by an acquisition period subdivided in separate runs of data taking, each run lasting for about one day.
The inclined campaign started in December 2014 and lasted for two weeks, while the vertical campaign started in February 2016 and lasted for six weeks.
A total of 594500 muons were collected.
A configuration scheme of the hodoscope geometry used for the experimental setup in both campaigns is shown in~\cref{fig:setup2}.

During the first campaign, the two RPCs were set at the opposite sides of the WCD to select inclined muons.
The whole of the top RPC and half of the bottom RPC were instrumented with 64 and 32 pads, respectively.\footnote{~The PREC acquisition system had a total of 104 channels.
Due to space constraints, we decided to install only four front-end boards in the bottom RPC.}
No calibration histograms from the WCD were recorded in this campaign, only the triggered data was taken.

WCD calibration data were, in turn, recorded during the acquisition with the vertical setup.
Both RPCs were positioned close to the central axis of the WCD to select vertical atmospheric muons passing through near the center.
Only 9 pads in the top RPC and 6 in the bottom RPC were connected to the data acquisition system to maximize the coincidence rate in the central region.
The narrowing of the active area was needed since the WCD acquisition imposes a large dead time, which would limit the readout of useful data when the whole hodoscope area is used for coincidence.

While the active area of the top RPC was covering the geometric center of the WCD, this was not the case for the active area of the bottom RPC.
The nearest pad of the bottom RPC was at a distance of ${\sim}20$\,cm from the center since a central pillar of the WCD support structure prevented us from moving it closer.
This is the main reason why the acquired muon trajectories in this vertical setup are not exactly centered within the WCD.

The background rate in each pad was continuously monitored and a pad was excluded from the trigger when its rate exceeded 1000\,s$^{-1}$.
Typical single-pad rates were of the order of a few hundred s$^{-1}$ for the top RPC and a few tens of s$^{-1}$ for the bottom RPC.
Such a difference is expected since most of the electromagnetic signal in the background cosmic ray beam is shielded by the water and does not reach the bottom RPC.
Moreover, since it is not in the shade, the top RPC is subject to higher temperature fluctuations, which lead to higher gas gain when temperature is higher.
Background events are removed from the data in the offline analysis.

\section{Simulation}
\label{SecSimulation}

A dedicated simulation was developed to assess the results obtained with the two experimental setups.
In a first step, a low-energy simulation of air showers provided an energy and zenith-angle dependent flux of background particles including atmospheric muons and other secondary particles at the ground level.
In a second step, particles were sampled from these distributions and were injected into a detailed detector simulation.
It should be noted that the accurate description of secondary particles at the ground depends on several volatile environmental quantities such as the solar modulation of the cosmic-ray intensity or the atmospheric conditions like temperature and pressure.
This results in numerous systematic uncertainties in the description of the background particle distributions.
Hence, only the main features of the data are expected to be reproduced by the simulations and not all of the fine details that might be observed in real conditions.

\subsection{Shower simulation}

The simulation of showers was performed using the \textsc{Corsika}~\cite{CORSIKA} framework.
Primary cosmic rays of different species were isotropically injected onto the top of the atmosphere with relative fluxes in the energy range between $10^{10}$ to $10^{15}$\,eV taken from Ref.~\cite{SimHernan}.
Here, the lower energy limit was set due to the geomagnetic-field cutoff.
The high-energy limit of the primary cosmic rays was chosen so that the corresponding flux then results in a negligible number of background particles in an experiment with the effective area of ${\sim}10$\,m$^2$.
Secondary shower particles in the showers simulations were recorded at an altitude of 1400\,m a.s.l., corresponding to the mean altitude of the Auger site.
The magnetic field of the Earth at the site was also used to define the primary cosmic-ray abundances and was taken into account in the simulation during the propagation of particles.
These simulations show that the low-energy secondary particles reaching the ground are mainly photons.
However, above an energy of a few GeV the population is entirely dominated by muons.

\begin{figure}[t]
\centering
\includegraphics[width=0.7\textwidth]{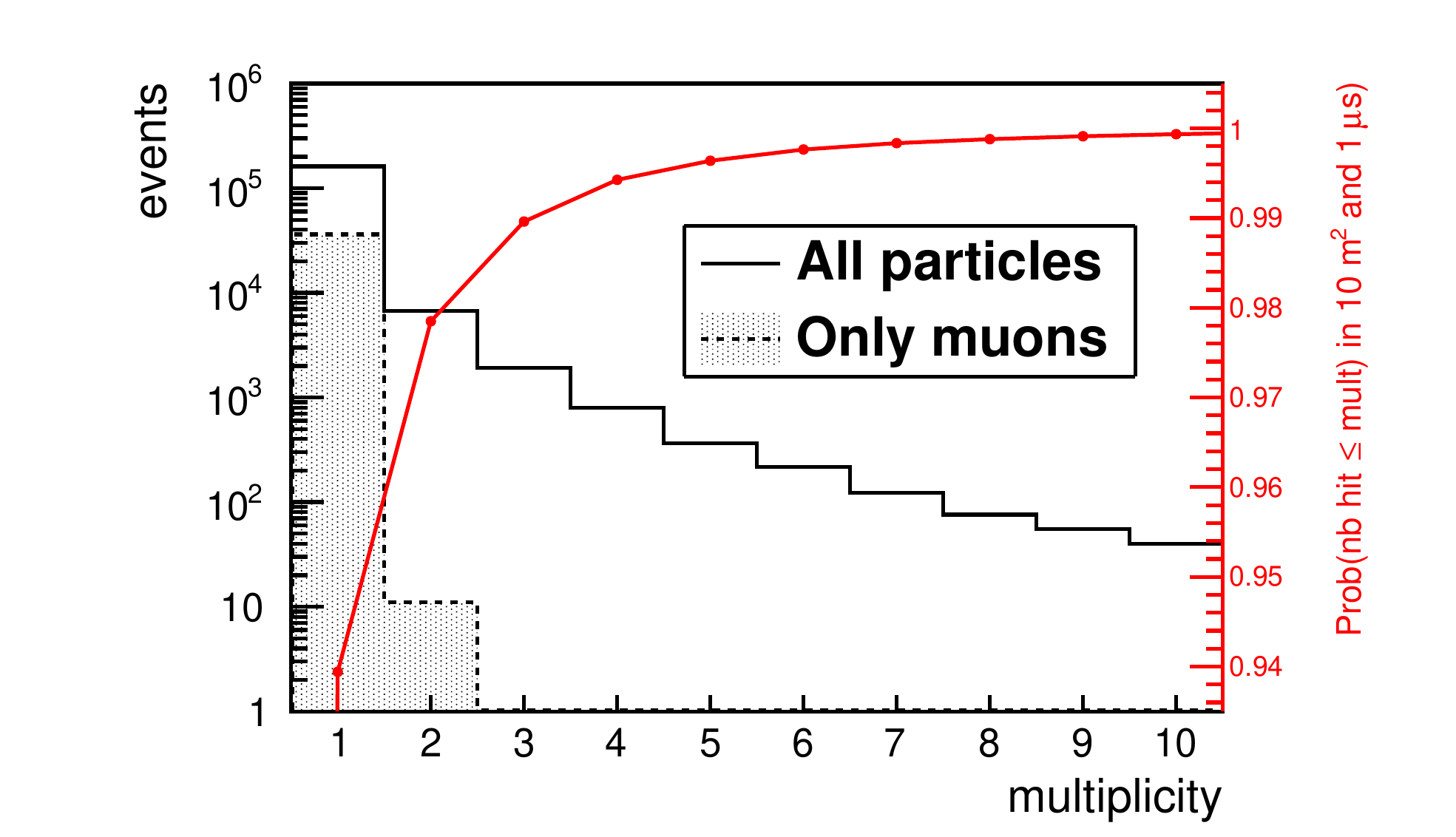}
\protect\caption{Histogram of the number of atmospheric particles reaching the ground at an altitude of 1400\,m a.s.l.\ in a detection area of 10\,m$^2$ during a time window of $1\,\upmu$s.
The shaded histogram corresponds to muons only.
The red curve shows the probability that the number of hits in a RPC is smaller or equal than the number of particles.}
\label{fig:AtmSimShowers}
\end{figure}

Using the distributions of background particles produced, the influence of multiple time-correlated events in the experimental apparatus, such as muon bundles, could be investigated.
The distribution of the number of particles crossing a detection area of 10\,m$^2$ (corresponding to the base area of the WCD) in a time window of $1\,\upmu$s (corresponding to the maximum time window for the RPC-coincidence trigger) is shown in~\cref{fig:AtmSimShowers}.
The probability of observing more than one atmospheric particle in this time window is less than 6\%.
Additionally, the hatched part of the distribution indicates that the probability of finding two muons in the WCD in the same time window is negligible.
From these results, combined with the selection criteria further described in the text, we conclude that a simulation of the SD gives an adequate comparison to the measured data when the atmospheric muons are injected individually.

\subsection{WCD simulation}
The simulation of the SD setup was performed using the Auger \Offline software framework~\cite{Offline}.
In~\Offline, the WCD response to entering particles is modeled using the \textsc{Geant4} software framework~\cite{Geant4-1,Geant4-2,Allison:2016lfl}.
Properties of a typical WCD, such as the geometry and all the materials composing the different components of the SD, are defined as an input to the simulation code.
To keep the simulation time within reasonable limits, only the physics processes that result in non-negligible response contributions were considered, namely:
\begin{itemize}
\item for \emph{photons:} photoelectric effect, Rayleigh/Compton scattering, and pair production (conversion);
\item for \emph{electrons/positrons:} multiple scattering, ionization, bremsstrahlung, Cherenkov emission, and annihilation (for positrons only);
\item for \emph{muons:} multiple scattering, ioniziation, bremsstrahlung, pair production, Cherenkov emission, delta-ray emissions, and muon decay;
\item for \emph{hadrons:} multiple scattering and ionization.
\end{itemize}

Each Cherenkov photon emitted in the wavelength range between 250 and 700\,nm is propagated within the water volume: its diffuse reflection is performed on the walls of the Tyvek container until it either reaches the photocathode of one of the three PMTs or it is absorbed in the water or Tyvek.
The key parameters governing this part of the simulation are the water absorption-length and the Tyvek reflectivity, which are both wavelength dependent.

When a photon reaches the photocathode of a PMT, the emission of the photoelectron is simulated according to the quantum efficiency of the PMT, which is also wavelength-dependent.
The current produced by the avalanche in the PMT dynodes is extracted both at the anode and at the last dynode, the latter signal being additionally amplified.
The quantum efficiency is here a key parameter.
More details on the simulation of the PMT can be found in~\cite{Creusot:2010xe}.
Then, the front-end electronics, including the amplification of the dynode signal, is simulated up to the digitization by subsequent 10\,bit fast analog-to-digital converters running at the sampling frequency of 40\,MHz, resulting in six FADC traces (two per PMT).

Accurately establishing the water absorption length, the Tyvek reflectivity, and the quantum efficiency of the PMTs for each of the WCDs turns out to be difficult.
The precise knowledge of the values of these wavelength-dependent parameters, which directly influence the number of photoelectrons produced by each PMT, is however not crucial for understanding the response of the detectors to showers.
The WCDs are continuously calibrated with atmospheric muons so that the measured signals from air showers are given in units of the equivalent charge produced in PMTs by a vertical muon crossing the middle of the tank, i.e.\ a vertical-equivalent muon (VEM) unit.
Consequently, in the WCD simulation we use average measured values for these parameters or their realistic estimates quoted in the literature.
The parameters, as well as their dependence on the wavelength, are given in \cref{SecSimulationParameters}.

The \Offline simulation includes the possibility of speeding up the simulation of the WCD.
Cherenkov photons are produced by charged relativistic particles traversing the water volume of the WCD.
Instead of simulating the propagation of these photons with the detailed \textsc{Geant4} stepper, the \emph{fast} option allows for custom wall-to-wall tracking where only the attenuation in the water and the diffusive scattering at the walls are treated.
With this option, the simulation typically runs five times faster, while preserving the number and directional distributions of the Cherenkov photons.

\subsection{RPC simulation}

The detailed description of the RPC structure, including the aluminum case, acrylic box, glass plates, and composition of the gas, was also implemented in the \textsc{Geant4} simulation.
The charged-particle track-lengths in the gas are recorded for subsequent processing.
A dedicated \Offline module is used to simulate the signal of the RPCs.
The generation of the signal in each pad is accounted for by using a parameterization of the RPC signal amplitude as a function of the particle inclination relative to the RPC.
This parametrization was obtained in laboratory measurements~\cite{RPCmeas} and takes into account the dependency of the charge produced by the RPC on the direction of the incoming particle.
Finally, the signal digitization using a simple threshold is also included in the RPC simulation chain.

With the approach described above, the tracking of particles through the WCD and the RPCs was performed in a seamless and consistent manner, enabling us to test directly the WCD simulation and, in particular, the WCD signal response to the passage of atmospheric muons.
Since the probability of having two background muons traversing the WCD at the same time is negligible and the contribution of high-energy showers can be effectively removed by imposing single hits in both RPCs, it is sufficient to simulate injections of only single atmospheric muons.
This avoids further complications in the analysis and makes the comparison between the simulation and data more reliable.
The simulated atmospheric muons were injected uniformly at the top RPC while preserving its arrival direction taken from the \textsc{Corsika} simulations.

All the simulation results presented in this paper have sufficiently high statistics and use the \emph{fast} simulation mode.
We ensured that the number of events in the simulation corresponded roughly to the amount of available data, i.e.\ of the order of $10^6$ events.
We also generated a smaller simulation sample using the full-\textsc{Geant4} mode with about 10\% of the statistics accumulated with the \emph{fast} simulation.
We find that both simulations give compatible results within the statistical uncertainties.

\section{Data reconstruction and performances}
\label{SecDataAnalysis}

\subsection{Charge reconstruction with the WCD}
\label{SecChargeWCD}

The WCD data provide a measurement of the signal charge associated with the muon detected by the hodoscope coincidence trigger.
To estimate the signal charge, we exploit the high-gain traces from the three PMTs, as well as their sum. 
We first evaluate the mean values of the baselines for each of the three channels. 
The trigger logic is such that traces are expected to have a signal with a maximum between time-bins 242 and 244. 
The baselines are thus evaluated from the portion of traces before the trigger region, where no signal is expected.

We then subtract the baselines from the FADC counts and search for signals in the traces. 
To avoid triggers generated spuriously by the electronics, only events that have the signal in the trigger region in all three PMT traces will be used in the following.
To identify the beginning and the end of the signal we adopt a threshold of 2.5\,ADC counts, which is high enough to remove any spurious effects from baseline noise. 
We take as start-time of the signal the first time-bin above such a threshold.
We define as stop-time the 10th time-bin after the baseline recovers from an undershoot to its nominal value. 
The charge of the signal is then obtained by integrating the baseline-subtracted ADC counts between the start- and stop-times. 
We have studied the dependence of the charge on the integration range, in particular on the upper limit of the integration. 
We have verified that the chosen criterion ensures the inclusion of the tail part of the signal while minimizing the effect of the baseline undershoot.
The uncertainty associated with this procedure is 1.5\,ADC counts. 
As the average charge due to a muon is about 180 in ADC-time units, this corresponds to ${\sim}0.01$\,VEM.

To convert the charge from the unit of ADC-time counts to the reference unit of VEM we use a calibration factor, which is extracted from the calibration histograms of the charge (see \cref{SecDAQTrigger}) and then scaled\footnote{~The procedure to calibrate the traces for the inclined data set, where no calibration data have been acquired, is addressed in \cref{SecResultsResponse}.} to the charge of a VEM, $Q_\text{VEM}$. 
We use in particular the position of the local maximum of the histogram, $Q^\text{peak}_\text{OD}$, which corresponds to the most probable charge deposited by omni-directional muons. 
This is indeed proportional to the equivalent charge of a vertical muon, i.e.\ 1\,VEM, through a scaling factor, $f_Q$, determined through a one-time measurement with a WCD equipped with a dedicated muon-telescope (see~\cite{NIM-A-Calib}).

\subsection{Trajectory reconstruction with the RPC}
\label{SecRecRPC}

The RPCs provide the reconstruction of the trajectory of the charged particles which traverse the hodoscope. 
To identify single muons, we select events that have only a \emph{single hit} in each of the two RPCs. 
With simulation, we have verified that this criterion removes most of the events related to showers in which more than one particle is crossing the top RPC. 
In this way also events are rejected in which a single atmospheric particle showers in the water, thus generating more than one hit in the bottom RPC. 
The simulation shows that the fraction of such events is less than 3\%.

In turn, the single-hit criterion does not reject events in which two uncorrelated atmospheric particles (see \cref{fig:AtmSimShowers}) or particles from the same shower pass through the hodoscope within the trigger time-window by chance. 
Nevertheless, such chance coincidences in fact represent less than $6\%$ of events.
Yet, some of the shower events produce signals large enough to saturate the high-gain PMT traces. 
We thus reject also events in which a saturation occurred in the WCD signal. 
An additional criterion that we use to remove background events from the data set is based on the ratio between the charge of the WCD signal and its maximum, the so-called ``Area over Peak'', $\text{AoP}$~\cite{Auger-NIM-A}. 
The muon signal is characterised by a fast rise followed by an exponential decay: on average $\text{AoP}\approx3.6$\,time bins (90\,ns). 
On the other hand, events triggered by radiofrequency noise have time bins with negative values, resulting in a much smaller value of AoP. 
We thus remove events from the data set when $\text{AoP}\leq1$\,time bins.

For the events selected as described above, we use the centers of the hit pads to reconstruct the muon trajectory. 
The active region of the RPC hodoscope allows us to reconstruct 54 different sets of muon trajectories in the vertical setup and 2048 sets in the inclined setup, corresponding to the number of pad combinations for top and bottom pairs. 
The RPC positions relative to the WCD have been measured to an accuracy of a few centimeters. 
The reconstructed entry point of the muon in the water volume of the WCD is ${\sim}1.3$\,m away from the top RPC, whereas the bottom RPC is very close to the WCD base. 
We define the muon impact parameter as the arithmetic mean between the WCD entry and exit points and use it to calculate the horizontal distance to the center of the WCD. 
We also reconstruct the length of the muon track inside the water, $\ell$, which is the relevant quantity for the amount of the produced Cherenkov light. 
Finally, we reconstruct the muon zenith angle as $\cos\theta=1.2\,\text{m}/\ell$ since with the selection criteria adopted we are not observing any clipping trajectories of the muon.
The corresponding angular resolution was obtained from simulations and is $1^\circ$.

\subsection{Zenith angle and charge distributions}
\label{SecChargeTraj}

After applying the selection criteria described in the two previous subsections, the data set of atmospheric muons traversing both the WCD and the RPC hodoscope includes 243961 (350539) vertical (inclined) events. 
The distributions of the reconstructed event parameters, zenith angle and charge, are compared below with those obtained for simulated events.

\begin{figure}[t]
\def\figh{0.41}
\centering
\includegraphics[height=\figh\textwidth]{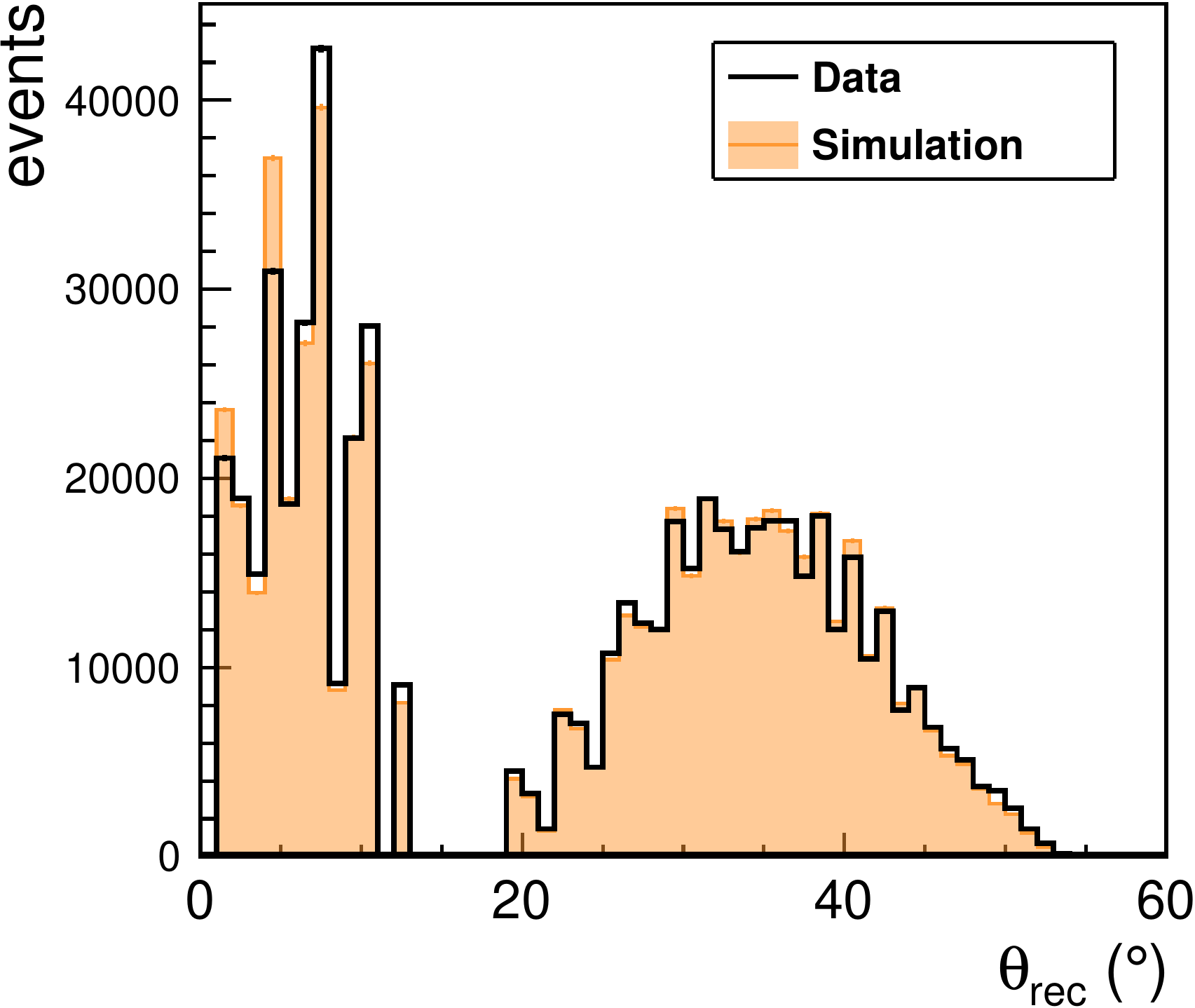}\hfill
\includegraphics[height=\figh\textwidth]{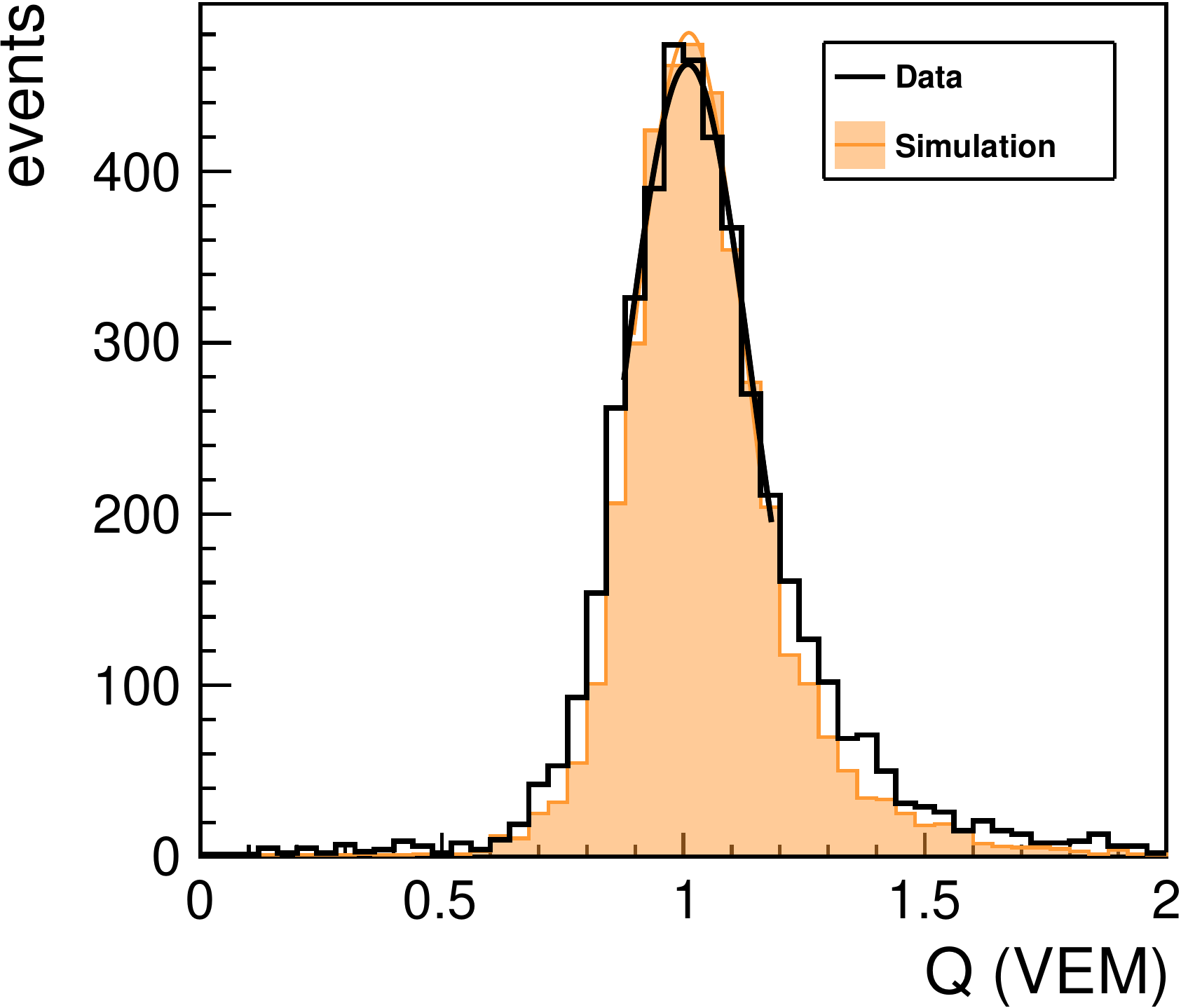}
\caption{\emph{Left:} Distribution of the zenith angles for measured (thick black line) and simulated (in filled orange) data.
The range of zenith angles covered by the vertical (inclined) configuration of the hodoscope is $0^\circ<\theta<14^\circ$ (${\sim}20^\circ<\theta<55^\circ$).
The use of the centers of the hit pads for the muon trajectory reconstruction produces the discretisation-aliasing effect.
\emph{Right:} An example charge distribution in VEM (thick black histogram) from the set of vertical events ($0^\circ<\theta<6.9^\circ$), with a Gaussian fit around the maximum (thin black line).
The corresponding distribution and the Gaussian fit for simulated events are shown as a filled orange histogram and as a thin red line, respectively.}
\label{fig:Geometry}
\end{figure}

In the left panel of \cref{fig:Geometry}, we present the distribution of zenith angles measured for the trajectories of the selected data (black histogram) compared to that obtained with simulated events (filled orange histogram). 
The width of the histogram bins of $1^\circ$ was chosen to correspond to the angular resolution.
The maximum zenith angle in the vertical setup (histograms on the left) is $14^\circ$ ($\sec\theta<1.03$), resulting in at most a 3\% deviation of the track length with respect to the vertical muons. 
The events in the inclined sample (histograms on the right) have zenith angles ranging from ${\sim}20^\circ$ to $55^\circ$.

The main features of the data histograms are well reproduced by the simulations, including the discretisation-aliasing effect, an artifact of the RPC granularity, which is observed in both measured and simulated events at exactly the same positions.
The main differences are observed for the vertical sample. 
Such differences are to a certain degree expected due to the limitations of the simulation to describe in detail the modulations of the atmospheric muon flux, as explained in \cref{SecSimulation}, affecting the distribution of zenith angles.
These modulations preferentially affect the softer part of the energy spectrum of the atmospheric muons and, therefore, the zenith-angle distribution of the vertical sample. 
The reproduction of the fine details of the observed histogram is, however, not needed for the purpose of this study.

For each set of muon trajectories, or each bin of muon track-lengths, we build the distributions of the charge for both measured and simulated data.
In the right panel of \cref{fig:Geometry} we show an example of such distributions, obtained from measured (black) and simulated (filled orange) data with zenith angles between $0^\circ$ and $6.9^\circ$ and the average reconstructed angle being $2.4^\circ$.
The position of the maximum of the distribution, $Q^\text{peak}$, is a suitable variable for the comparison of the two distributions in terms of response to muons.
The background, be it due to electromagnetic particles, small showers, or spurious triggers, indeed mostly only affects the tails of the distribution.
We first determine an approximate position of the maximum by means of a polynomial fit.
We then fit the 1-sigma range around the determined maxima with a normal distribution.
The fits are shown in \cref{fig:Geometry} as thick lines, with measured (black) and simulated (orange) data.
It is found that the maxima for measurements and simulations are well in agreement, being at 1.010 and 1.011\,VEM, respectively.
The uncertainty is ${\sim}0.01$\,VEM, which corresponds to ${\sim}1\%$ of the signal charge.
A similar agreement is found for all the different sets of muon trajectories.
The slight difference between the width of the distributions is explained by the different AoP of
this specific WCD with respect to the simulation.

\section{Response of the WCD to muons: data-simulation comparison}
\label{SecResultsResponse}

In the previous section we have shown how the distributions of the basic observables of the hodoscope, namely the direction (reconstructed by the RPCs) and the amplitude of the signal (recorded in the WCD) are well reproduced by simulations. 
This preliminary validation grants us the possibility to study the response of the WCD to muons in more detail, even down to the level of the response of individual PMTs. 
To this aim, we investigate the behavior of the signal charge as a function of the zenith angle, i.e.\ of the muon track-length $\ell$ in the WCD.

\begin{figure}[t]
\def\figh{0.41}
\centering
\includegraphics[height=\figh\textwidth]{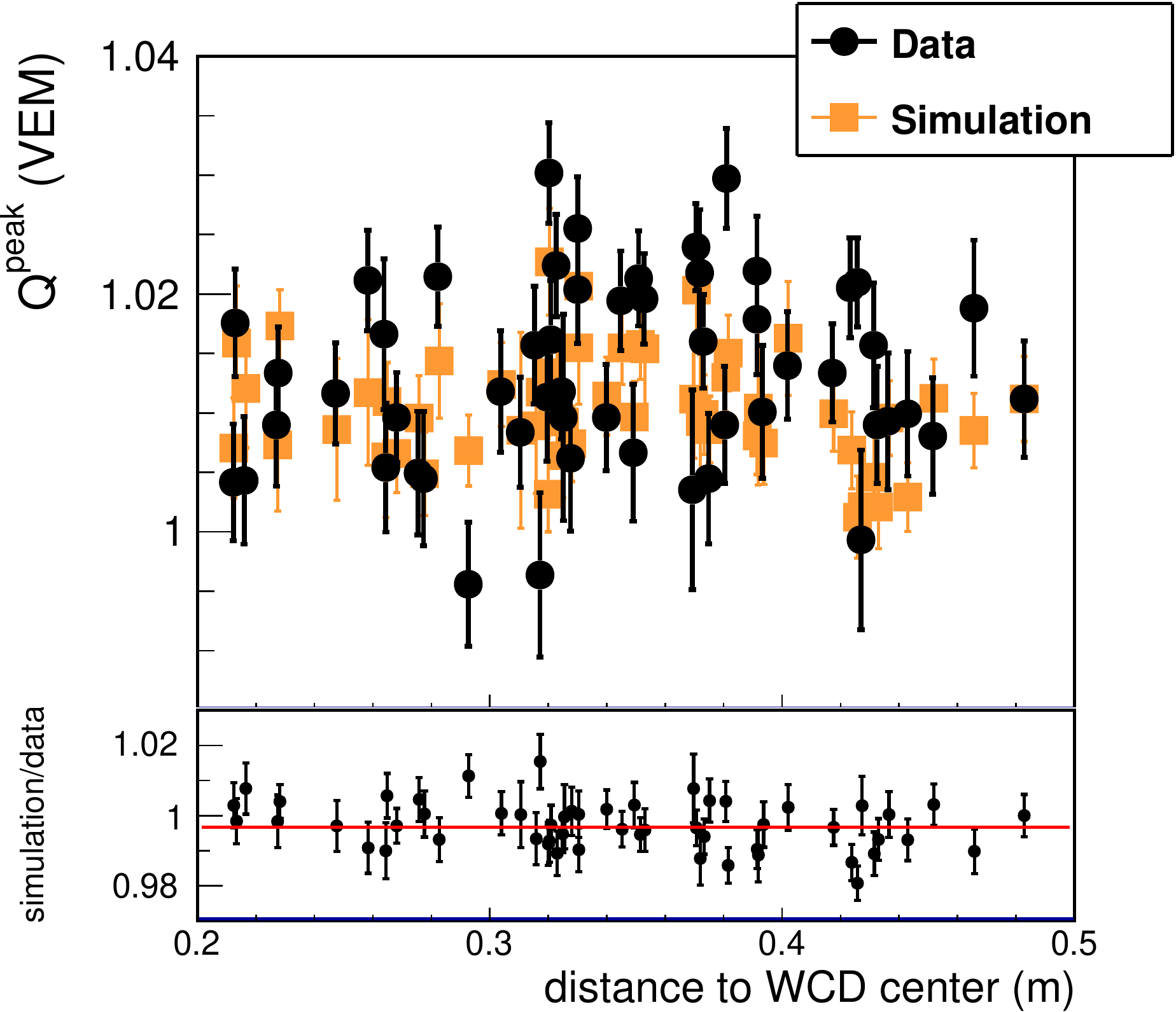}\hfill
\includegraphics[height=\figh\textwidth]{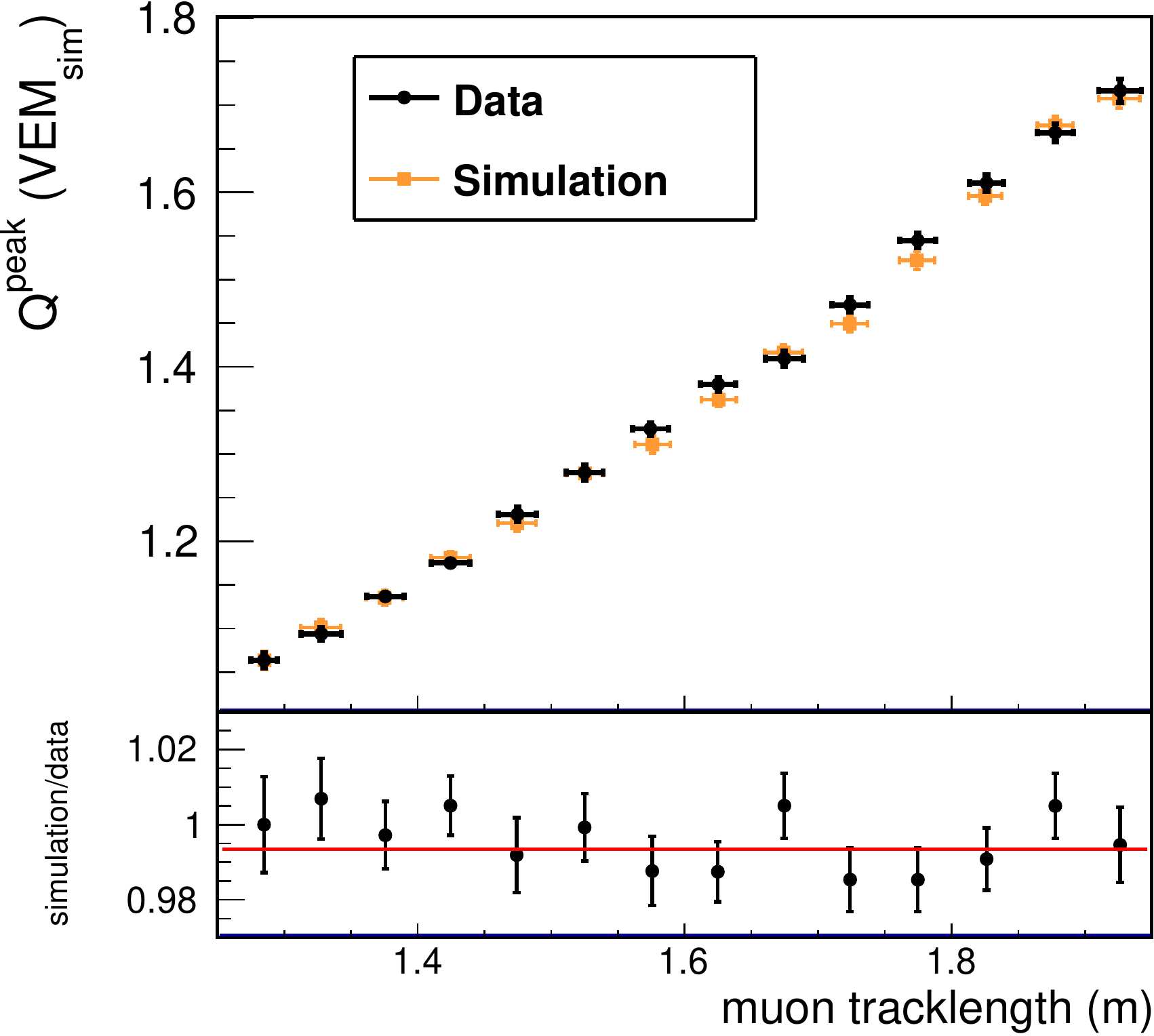}
\caption{\emph{Left:} Vertical events; fitted position of the maximum of the charge distribution in VEM as a function of the impact distance to the WCD center.
\emph{Right:} Inclined events; fitted position of the maximum of the charge distribution in VEM as a function of the track-length.
In both figures the charge is obtained from the sum of the three PMTs.}
\label{fig:DataVsOfflinePMTSum}
\end{figure}

For the near-vertical muons ($\ell=1.2$\,m), we have shown above (see \cref{fig:Geometry}, right) that the associated charge -- defined as the position of the maximum of the charge distribution -- is, as expected, 1.01\,VEM. 
Here, we further develop the study by examining the distributions of the charge associated with vertical muons as a function of the impact distance from the WCD center. 
As explained in \cref{SecExperimentalSetup}, due to the geometry of the hodoscope, the trajectory of vertical muons is in fact not centered in the WCD.
A lack of dependence on distance to the center is indeed expected: this is due to the fact that while individual PMTs can see signals smaller or larger than the others, depending on the distance to the passing muon, such an asymmetry cancels to the first order when the sum of their signals is used. 
In the left panel of \cref{fig:DataVsOfflinePMTSum} we show the positions of the fitted maxima of the distributions of the charges (black dots), obtained from the sum of the three PMT signals, as a function of the distance to the WCD center. 
The error bars correspond to the uncertainties of the fit. 
One can see that there is no dependence on the distance, as observed in the simulations (orange squares): the bottom plot shows that the ratio between the measurements and simulations is within 1\%, i.e.\ at the level of the signal uncertainty, thus giving a further verification of the accuracy of the simulation.

To extend the study to $\ell>1.2$\,m, we exploit the data collected by the hodoscope in the inclined configuration. 
The result is shown in the right panel of \cref{fig:DataVsOfflinePMTSum}, where we compare the fitted position of the maxima of the charge distributions\footnote{~Note that as calibration histograms are not available for this configuration, the conversion of ADC-time counts in units of VEM is performed using the position of the maximum of the charge distribution of simulated events, for which $1.25<\ell/\text{m}<1.3$.
The conversion constant is therefore $Q^\text{peak}_\text{sim}$/$Q^\text{peak}_\text{data}$ evaluated at this track-length bin.
This is the most vertical sample observed with the inclined setup: we denote the corresponding $Q^\text{peak}$ in $\text{VEM}_\text{sim}$ units.} determined with measurements (black dots) and simulations (orange squares), as a function of $\ell$. 
The expected increase of the charge as $\ell$ increases is consistently observed in data and simulations: as shown in the bottom inset, also for inclined events the agreement is at the level of 1\%.

\begin{figure}[t]
\def\figh{0.41}
\centering
\includegraphics[height=\figh\textwidth]{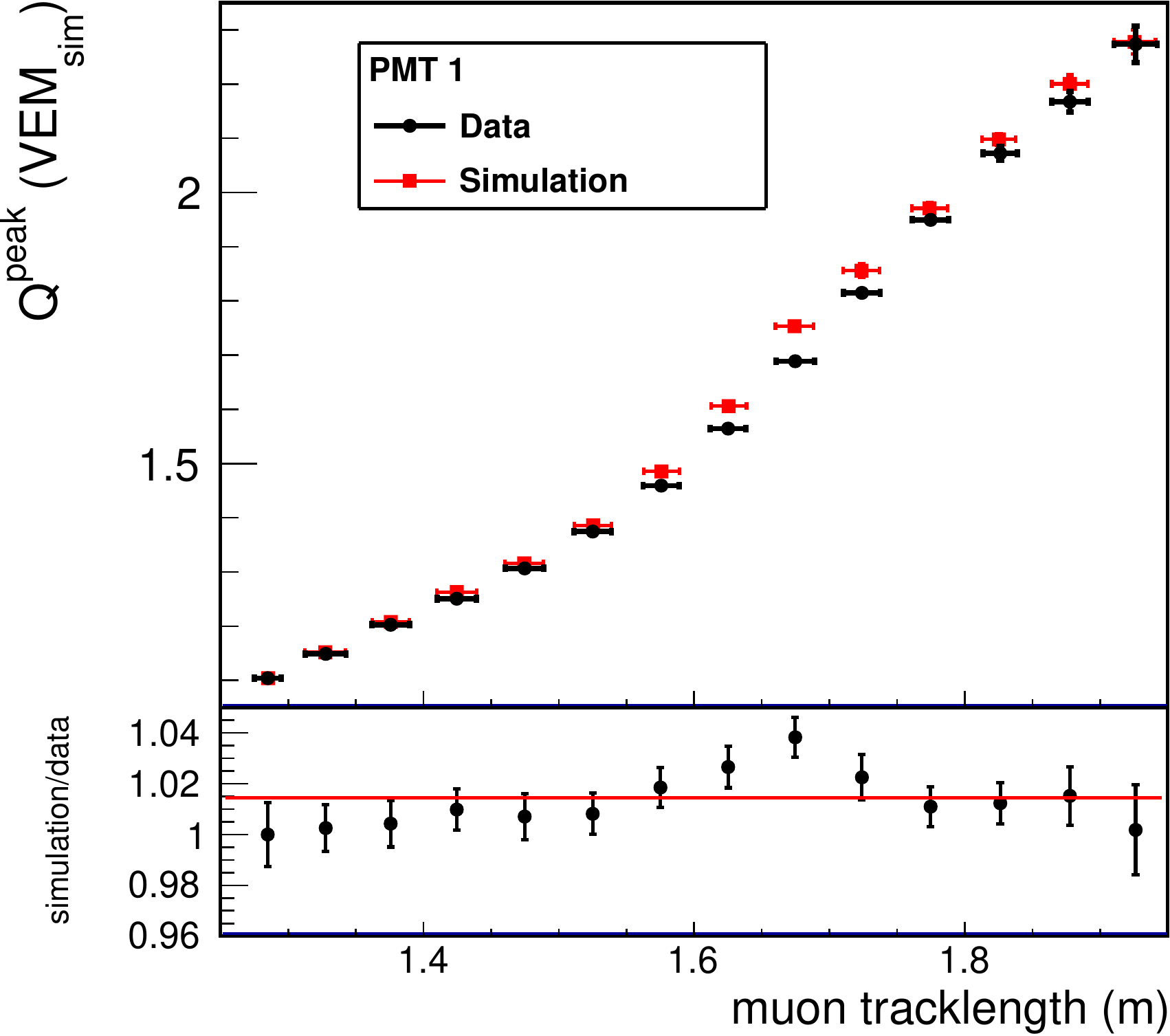}\hfill
\includegraphics[height=\figh\textwidth]{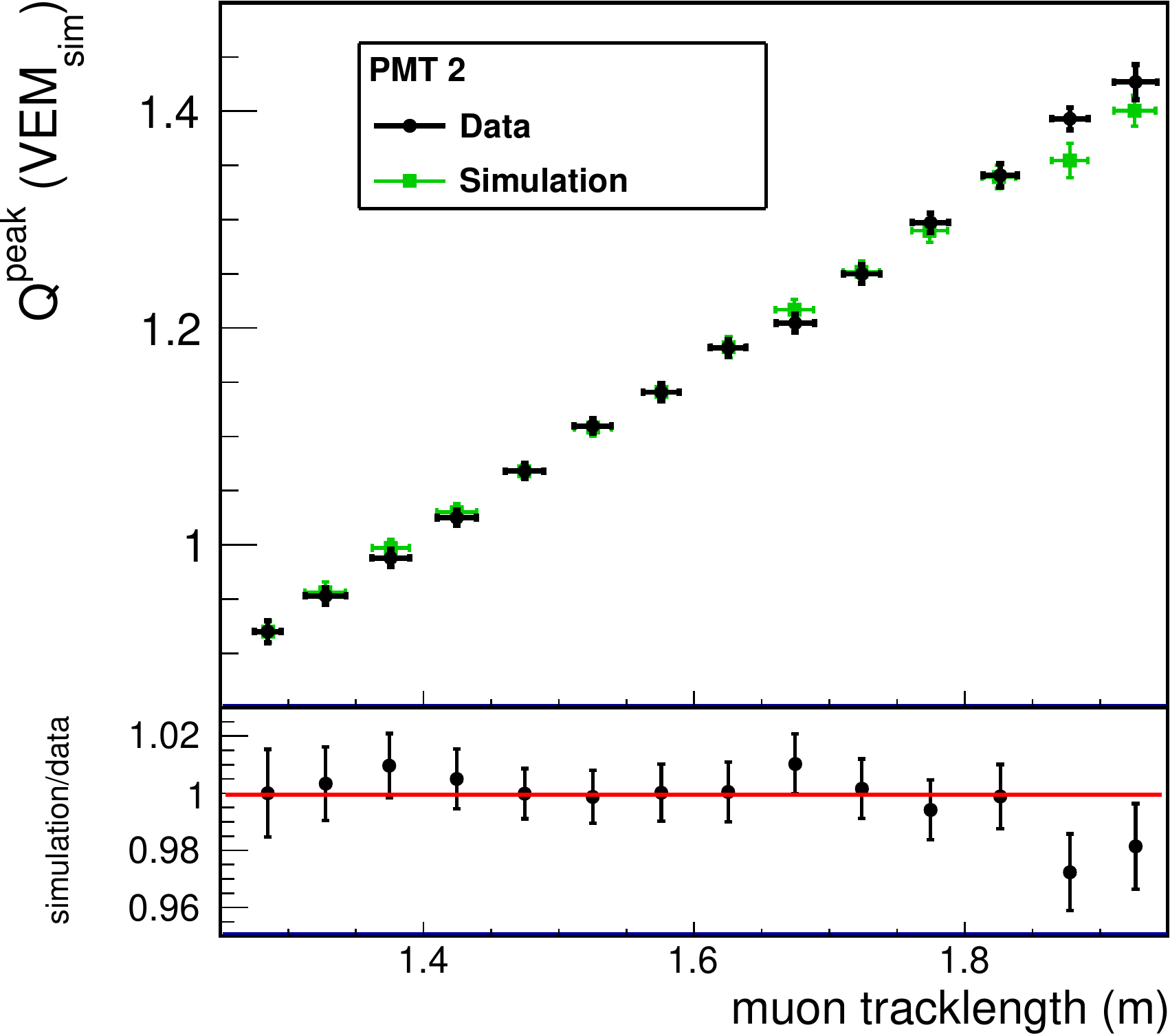}
\\
\includegraphics[height=\figh\textwidth]{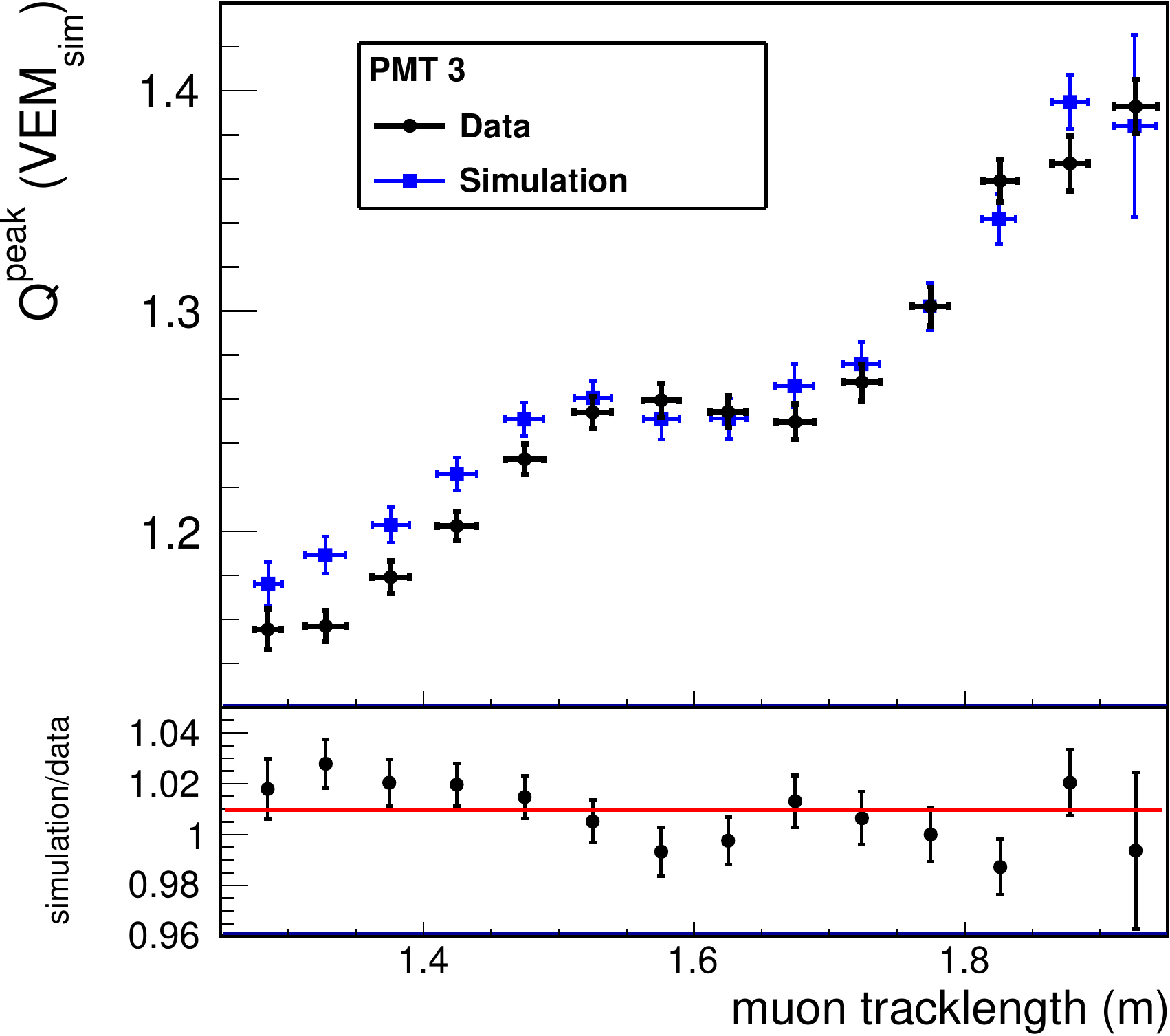}
\caption{Position of the peak of the charge distribution as a function of the muon track-length in the WCD water, for individual PMTs.
The colored points represent the simulations, while the black points correspond to measurements.}
\label{fig:SvdLPMTs}
\end{figure}

A finer verification of the simulation is viable by studying the response of individual PMTs as a function of the track-length, given that the geometry of the inclined setup is expected to induce differences between the three PMTs (see \cref{fig:setup2}, right). 
The behavior of the maximum of the charge distribution as a function of $\ell$ is shown in \cref{fig:SvdLPMTs}, for PMT\,1, PMT\,2, and PMT\,3, as left, right, and bottom panels, respectively.

Clear differences are observed among the PMTs and the uniqueness behavior for each is consistent between the measurements and simulations. 
The muon trajectories reach the bottom of the WCD closer to the position of PMT\,1 as the zenith angle increases.
A larger amount of Cherenkov light due to the first reflection from the liner is expected as the zenith angle increases, and also from ``direct'' Cherenkov light (i.e.\ photons with no reflection). 
The change of slope for PMT\,1 at ${\sim}1.6$\,m, seen in measurements and in simulations is well explained by the increasing contribution of the non-diffused light for the longest track-lengths. 
In turn, PMT\,2 is located opposite to the muon trajectories selected by the hodoscope so that it is reached mostly by well-diffused Cherenkov light. 
The observed dependence of the VEM charge with the muon track-length is therefore linear. 
Finally, for PMT\,3, ``direct'' Cherenkov light  or Cherenkov light generated in the PMT glass is expected for the shortest track-lengths.\footnote{~For the PMT\,3, the ADC conversion is done using the $Q^\text{peak}$ determined with simulated events with $1.75<\ell/\text{m}<1.8$, for which the signal is dominated by diffuse light.}
In these particular short tracks, such components of Cherenkov light vary rapidly with the geometry and lead to the observed non-linearity up to 1.7\,m. Above, the linear behavior expected from well-diffused Cherenkov light is observed.

Also at the level of individual PMTs, the response of the WCD is well reproduced by simulations: the insets at the bottom of the three figures show that their ratios lie to within a few percent around 1, with a maximum deviation of 4\% for the PMTs which have the most peculiar geometry in the considered configuration.

\section{Scaling factor for the VEM calibration}
\label{SecResultsVEM}

\begin{figure}[t]
\centering
\includegraphics[width=0.7\textwidth]{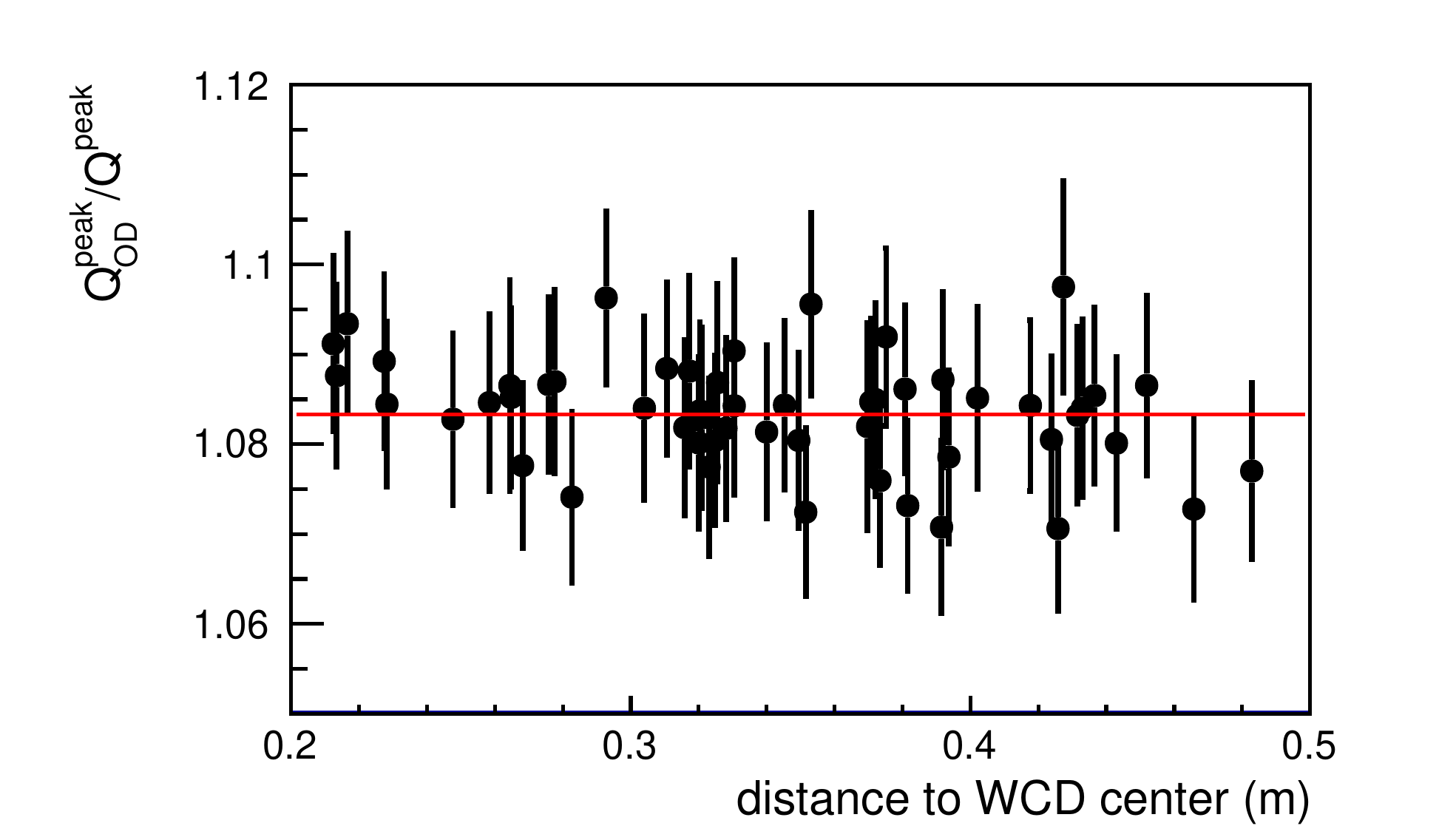}
\caption{The ratio $Q^\text{peak}_\text{OD}/Q^\text{peak}$ as a function of the impact distance to the WCD center, determined by the sum of the PMT charges.}
\label{fig:CFchargePMTsum}
\end{figure}

The directional capability of the hodoscope and the good understanding of the response of the system motivated us to use the data also to perform a new measurement of the scaling factor $f_Q$ of the WCD calibration between the average charge due to omni-directional and vertical muons, i.e.\ $f_Q=Q^\text{peak}_\text{OD}/Q_\text{VEM}$. 
As mentioned in \cref{SecChargeWCD}, the one-time measurement of $f_Q$ was performed at the beginning of the operation of the Observatory~\cite{ProcCalib2005} by means of a scintillator-based muon telescope built around the same test WCD, which enabled selection of centered vertical muons.

In turn, and as already remarked in the previous section, the data sample of the RPC hodoscope does not include vertical muons crossing the WCD center. 
The range of distances from the center is in fact between 20 cm and 50 cm (see \cref{fig:DataVsOfflinePMTSum}, left), which we consider large enough to determine the behavior of the ratio $Q^\text{peak}_\text{OD}/Q^\text{peak}$ as a function of the distance, so as to extrapolate from it the value corresponding to central muons.
As for the studies shown in previous sections, $Q^\text{peak}$ is the position of the fitted maximum of the charge distribution for each set of the non-centered vertical data, extrapolated to the center of the WCD.
The small deviations of the trajectories from the vertical are taken into account by normalizing $Q^\text{peak}$ to the tracklength such that they resemble the equivalent vertical trajectory.

In \cref{fig:CFchargePMTsum}, we show $Q^\text{peak}_\text{OD}/Q^\text{peak}$ as a function of the  distance to the WCD center.
As one can see, the ratio is constant over the range considered, as expected from simulations and verified by finding a slope statistically compatible with zero. This fact allows us to determine $f_Q$ from the error-weighted average over the distances.
The value obtained, $1.08\pm0.01$, is represented in the figure by the red line.
The uncertainty is obtained by adding in quadrature the systematic uncertainty of the charge determination (see \cref{SecDataAnalysis}) and its statistical uncertainty.

The value obtained with the new measurement is in agreement with that found 15 years ago ($1.09$, see \cite{NIM-A-Calib}), both measurements being performed around the local summer time, thus allowing us to verify the long-term stability of this ingredient of the WCD calibration procedure.

\section{Summary and conclusions}
\label{SecConclusions}

The understanding of the response of the Auger water-Cherenkov detectors (WCD) to muons is of prime importance, not only because muons serve as a convenient means of detector calibration, but also because they are messengers of the hadronic interactions that drive the development of extensive air showers in the atmosphere.

With the aim of gaining such understanding, we have designed and deployed a hodoscope based on resistive-plate chambers (RPCs).
Thanks to the excellent positional and directional performance of the hodoscope, it has allowed us to select single muons and reconstruct their direction with $1^\circ$ accuracy over a large range of zenith angles between $0^\circ$ and $55^\circ$.
The measurement of charge of the corresponding signal in the WCD has enabled us to study its response as a function of the zenith angle.

We have then compared the measured data with the expectations from a simulation, which has been implemented by using \textsc{Corsika} to simulate showers and \textsc{Geant4} to accurately model the response of WCDs and the RPCs.
We have found that, down to the level of the response of individual PMTs to muons, the agreement between the data and expectations is at a level of 2\%.

We have also taken advantage of the most vertical data of the hodoscope to verify one crucial element of the calibration chain, namely the scaling factor of the signal charge between omni-directional and vertical muons.
The value we obtained is in good agreement with the value measured 15 years ago, at the very beginning of the operation of the Pierre Auger Observatory.

In conclusion, the level of the agreement of the simulation with the presented measurements validates with high accuracy the Auger \Offline simulation of the WCD response to muons.
Furthermore, the updated measurement of the scaling factor used in the WCD calibration shows no evidence of ageing effects.

\appendix

\section{Surface Detector simulation parameters}
\label{SecSimulationParameters}

The amount of Cherenkov light produced by the particles entering a WCD of the surface detector and that reach the PMTs mainly depends on the characteristics of the water and the inner-wall lining made of Tyvek.
The first key parameter is thus the water absorption length, which is a function of the wavelength of the photons.
For its wavelength dependency we adopt the measurements made with pure water~\cite{Water:1981}.
The next major input into the simulation is the Tyvek reflectivity featuring a strong diffusive and a weaker specular component.
The latter is chosen to be constant and at the level of 20\%.
The diffusive component of the reflectivity is also a function of the photon wavelength~\cite{Tyvek:1999}.

\begin{figure}[t]
\def\figh{0.31}
\centering
\includegraphics[height=\figh\textwidth]{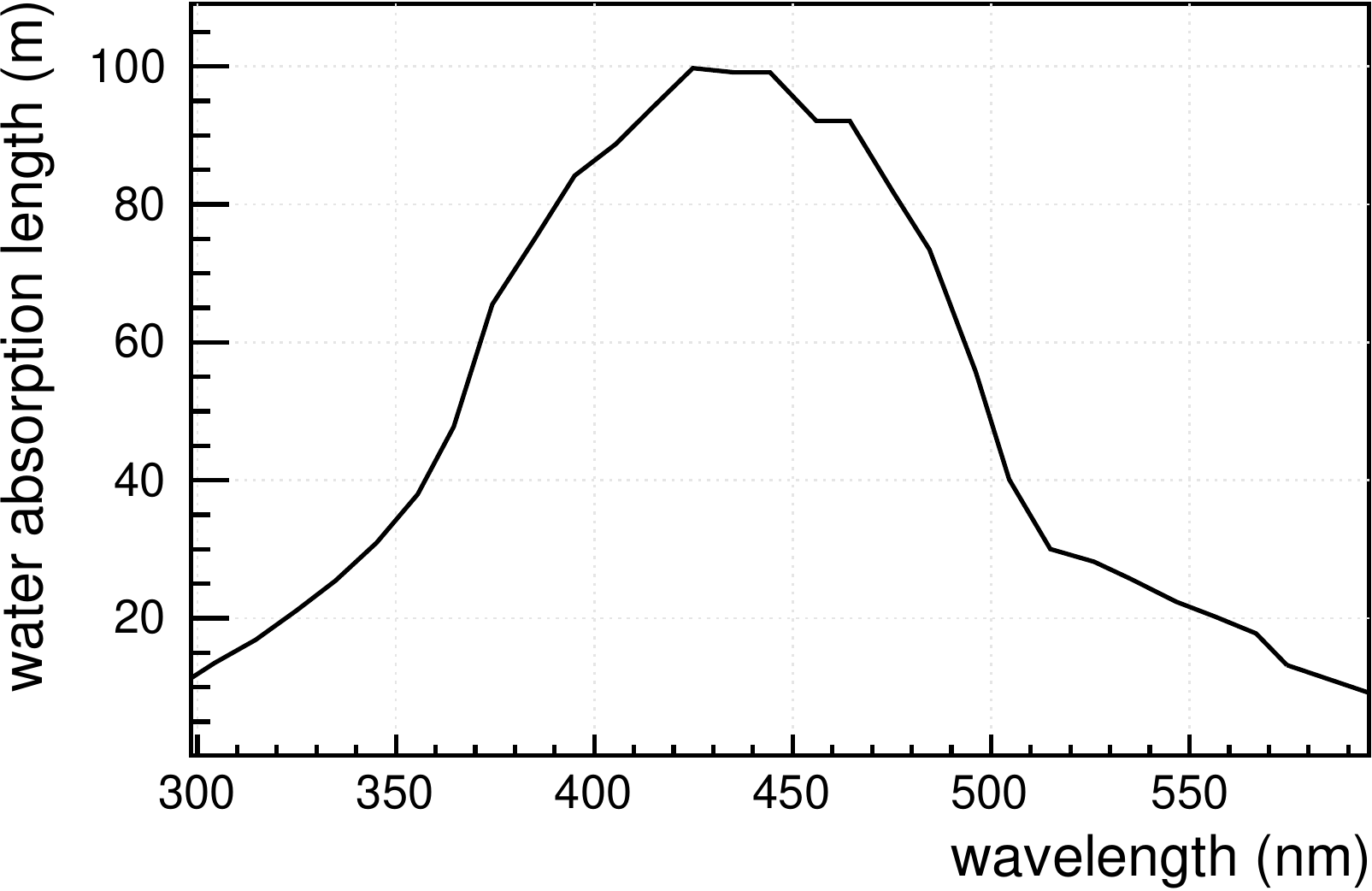}\hfill
\includegraphics[height=\figh\textwidth]{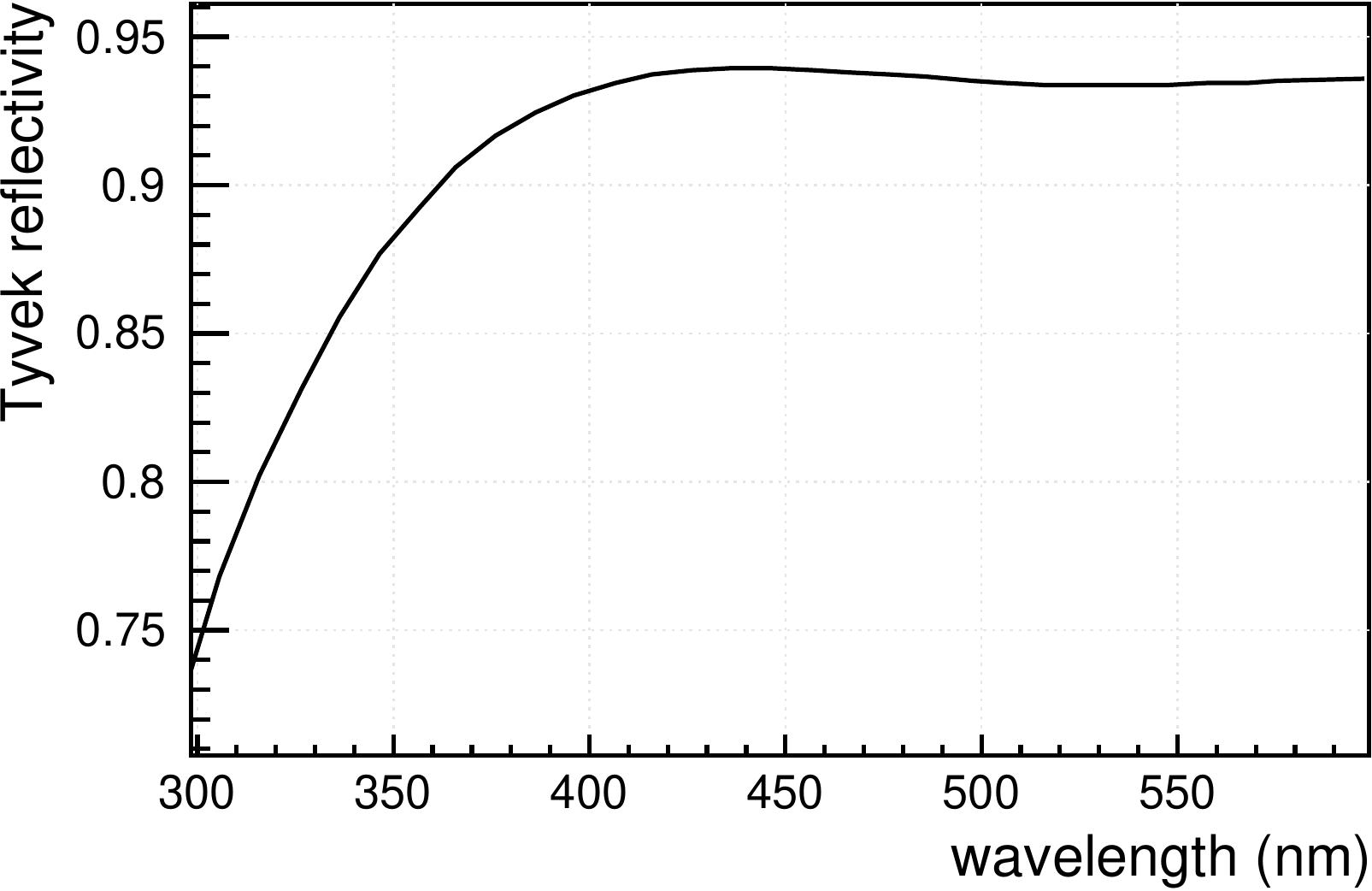}
\caption{\emph{Left:} Water absorption length as a function of wavelength.
\emph{Right:} Tyvek reflectivity as a function of wavelength.}
\label{fig:water-tyvek}
\end{figure}

The combination of these parameters directly influences the time spent by the Cherenkov photons in the water volume of the WCD before they are (i) absorbed or they (ii) reach the photocathode of one of the 3 PMTs.
The time can be measured in each WCD by observation of the decay time of pulses produced by muons.
The maximum values of the water absorption-length and the Tyvek reflectivity have been tuned so that they reproduce the decay time observed in a typical WCD.
While the 1660 WCDs in the SD array have slightly different behaviors, for the simulation we, nevertheless, chose only one set of parameters which represents a fair average of the measured values.
The maximum value for the water absorption-length has been set to 100\,m while the maximum Tyvek reflectivity is set to 94.0\%.
These two parameters are shown in the left and right panels of \cref{fig:water-tyvek}, respectively.

\begin{figure}[t]
\centering
\includegraphics[width=0.5\textwidth]{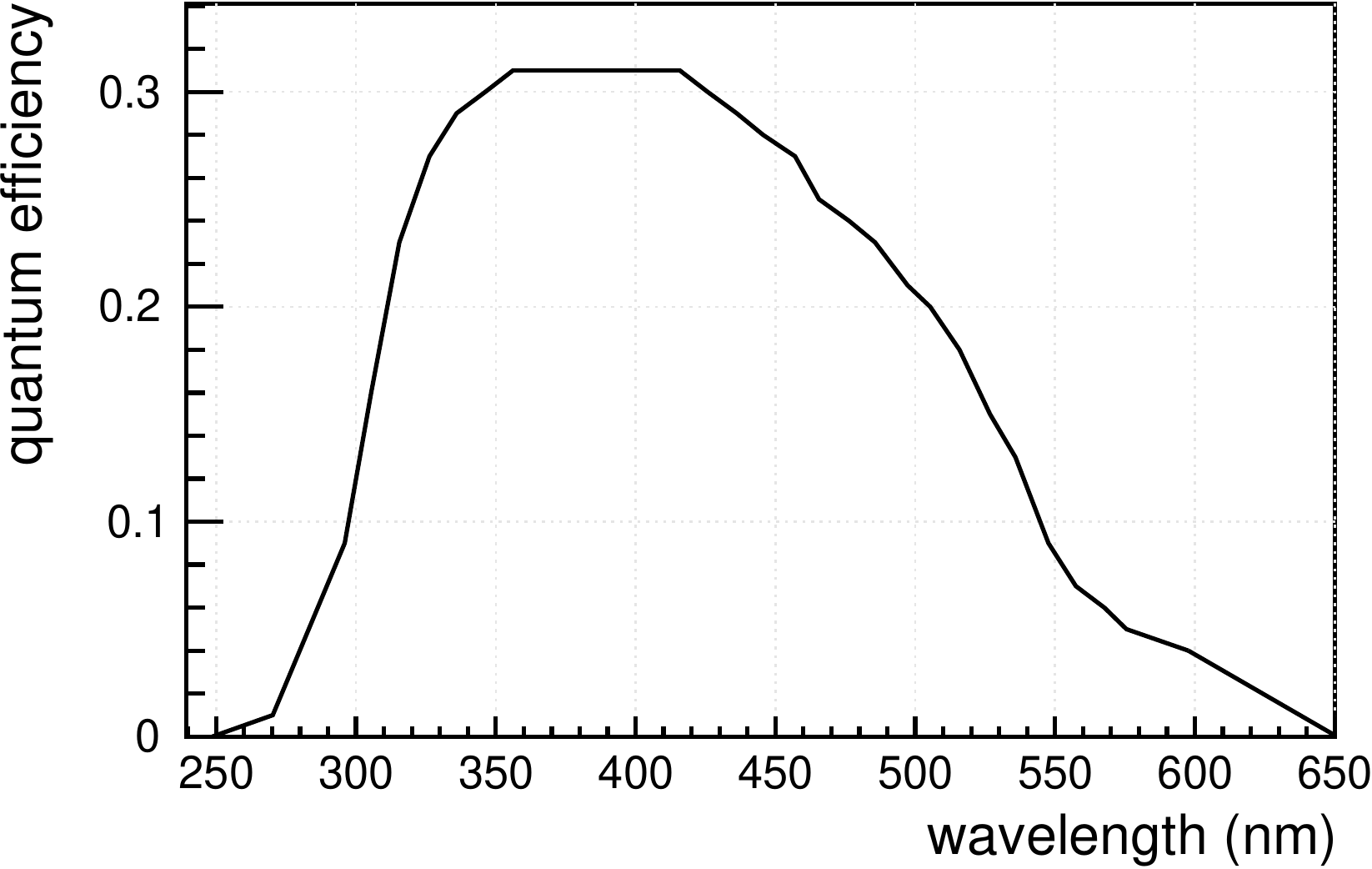}
\caption{PMT quantum efficiency as a function of wavelength.}
\label{fig:qe}
\end{figure}

The properties of the XP\,1805 photomultiplier tubes are also driving the response of the WCDs to particles.
We again use only constant parameters corresponding to their averages over all the WCDs.
The average size of the active photocathode surface of the PMT was fixed to 500\,cm$^2$.
The quantum efficiency of the PMT has a functional dependence on the photon wavelength as provided by the PMT manufacturer Photonis.
We modified it to take into account its angular dependence~\cite{Creusot:2010xe}, the result of which is shown in \cref{fig:qe}.


\section*{Acknowledgments}

\begin{sloppypar}
The successful installation, commissioning, and operation of the Pierre
Auger Observatory would not have been possible without the strong
commitment and effort from the technical and administrative staff in
Malarg\"ue. We are very grateful to the following agencies and
organizations for financial support:
\end{sloppypar}

\begin{sloppypar}
Argentina -- Comisi\'on Nacional de Energ\'\i{}a At\'omica; Agencia Nacional de
Promoci\'on Cient\'\i{}fica y Tecnol\'ogica (ANPCyT); Consejo Nacional de
Investigaciones Cient\'\i{}ficas y T\'ecnicas (CONICET); Gobierno de la
Provincia de Mendoza; Municipalidad de Malarg\"ue; NDM Holdings and Valle
Las Le\~nas; in gratitude for their continuing cooperation over land
access; Australia -- the Australian Research Council; Brazil -- Conselho
Nacional de Desenvolvimento Cient\'\i{}fico e Tecnol\'ogico (CNPq);
Financiadora de Estudos e Projetos (FINEP); Funda\c{c}\~ao de Amparo \`a
Pesquisa do Estado de Rio de Janeiro (FAPERJ); S\~ao Paulo Research
Foundation (FAPESP) Grants No.~2019/10151-2, No.~2010/07359-6 and
No.~1999/05404-3; Minist\'erio da Ci\^encia, Tecnologia, Inova\c{c}\~oes e
Comunica\c{c}\~oes (MCTIC); Czech Republic -- Grant No.~MSMT CR LTT18004,
LM2015038, LM2018102, CZ.02.1.01/0.0/0.0/16{\textunderscore}013/0001402,
CZ.02.1.01/0.0/0.0/18{\textunderscore}046/0016010 and
CZ.02.1.01/0.0/0.0/17{\textunderscore}049/0008422; France -- Centre de Calcul
IN2P3/CNRS; Centre National de la Recherche Scientifique (CNRS); Conseil
R\'egional Ile-de-France; D\'epartement Physique Nucl\'eaire et Corpusculaire
(PNC-IN2P3/CNRS); D\'epartement Sciences de l'Univers (SDU-INSU/CNRS);
Institut Lagrange de Paris (ILP) Grant No.~LABEX ANR-10-LABX-63 within
the Investissements d'Avenir Programme Grant No.~ANR-11-IDEX-0004-02;
Germany -- Bundesministerium f\"ur Bildung und Forschung (BMBF); Deutsche
Forschungsgemeinschaft (DFG); Finanzministerium Baden-W\"urttemberg;
Helmholtz Alliance for Astroparticle Physics (HAP);
Helmholtz-Gemeinschaft Deutscher Forschungszentren (HGF); Ministerium
f\"ur Innovation, Wissenschaft und Forschung des Landes
Nordrhein-Westfalen; Ministerium f\"ur Wissenschaft, Forschung und Kunst
des Landes Baden-W\"urttemberg; Italy -- Istituto Nazionale di Fisica
Nucleare (INFN); Istituto Nazionale di Astrofisica (INAF); Ministero
dell'Istruzione, dell'Universit\'a e della Ricerca (MIUR); CETEMPS Center
of Excellence; Ministero degli Affari Esteri (MAE); M\'exico -- Consejo
Nacional de Ciencia y Tecnolog\'\i{}a (CONACYT) No.~167733; Universidad
Nacional Aut\'onoma de M\'exico (UNAM); PAPIIT DGAPA-UNAM; The Netherlands
-- Ministry of Education, Culture and Science; Netherlands Organisation
for Scientific Research (NWO); Dutch national e-infrastructure with the
support of SURF Cooperative; Poland -Ministry of Science and Higher
Education, grant No.~DIR/WK/2018/11; National Science Centre, Grants
No.~2013/08/M/ST9/00322, No.~2016/23/B/ST9/01635 and No.~HARMONIA
5--2013/10/M/ST9/00062, UMO-2016/22/M/ST9/00198; Portugal -- Portuguese
national funds and FEDER funds within Programa Operacional Factores de
Competitividade through Funda\c{c}\~ao para a Ci\^encia e a Tecnologia
(COMPETE); Romania -- Romanian Ministry of Education and Research, the
Program Nucleu within MCI (PN19150201/16N/2019 and PN19060102) and
project PN-III-P1-1.2-PCCDI-2017-0839/19PCCDI/2018 within PNCDI III;
Slovenia -- Slovenian Research Agency, grants P1-0031, P1-0385, I0-0033,
N1-0111; Spain -- Ministerio de Econom\'\i{}a, Industria y Competitividad
(FPA2017-85114-P and FPA2017-85197-P), Xunta de Galicia (ED431C
2017/07), Junta de Andaluc\'\i{}a (SOMM17/6104/UGR), Feder Funds, RENATA Red
Nacional Tem\'atica de Astropart\'\i{}culas (FPA2015-68783-REDT) and Mar\'\i{}a de
Maeztu Unit of Excellence (MDM-2016-0692); USA -- Department of Energy,
Contracts No.~DE-AC02-07CH11359, No.~DE-FR02-04ER41300,
No.~DE-FG02-99ER41107 and No.~DE-SC0011689; National Science Foundation,
Grant No.~0450696; The Grainger Foundation; Marie Curie-IRSES/EPLANET;
European Particle Physics Latin American Network; and UNESCO.
\end{sloppypar}

\bibliographystyle{JHEP}
\bibliography{References}

\providecommand{\href}[2]{#2}\begingroup\raggedright\begin{thebibliography}{10}

\bibitem{Auger-NIM-A}
{\scshape Pierre Auger} collaboration, \emph{{The Pierre Auger Cosmic Ray
  Observatory}}, \href{https://doi.org/10.1016/j.nima.2015.06.058}{\emph{Nucl.\
  Instrum.\ Meth.\ A} {\bfseries 798} (2015) 172}
  [\href{https://arxiv.org/abs/1502.01323}{{\ttfamily 1502.01323}}].

\bibitem{Cazon2019}
L.~Cazon, \emph{{Probing High-Energy Hadronic Interactions with Extensive Air
  Showers}}, \href{https://doi.org/10.22323/1.358.0005}{\emph{PoS} {\bfseries
  ICRC2019} (2020) 005} [\href{https://arxiv.org/abs/1909.02962}{{\ttfamily
  1909.02962}}].

\bibitem{HASPRD2015}
{\scshape Pierre Auger} collaboration, \emph{{Muons in Air Showers at the
  Pierre Auger Observatory: Mean Number in Highly Inclined Events}},
  \href{https://doi.org/10.1103/PhysRevD.91.032003}{\emph{Phys.\ Rev.\ D}
  {\bfseries 91} (2015) 032003}
  [\href{https://arxiv.org/abs/1408.1421}{{\ttfamily 1408.1421}}].

\bibitem{GlennysPRL2016}
{\scshape Pierre Auger} collaboration, \emph{{Testing Hadronic Interactions at
  Ultrahigh Energies with Air Showers Measured by the Pierre Auger
  Observatory}},
  \href{https://doi.org/10.1103/PhysRevLett.117.192001}{\emph{Phys.\ Rev.\
  Lett.} {\bfseries 117} (2016) 192001}
  [\href{https://arxiv.org/abs/1610.08509}{{\ttfamily 1610.08509}}].

\bibitem{NIM-A-Calib}
{\scshape Pierre Auger} collaboration, \emph{{Calibration of the surface array
  of the Pierre Auger Observatory}},
  \href{https://doi.org/10.1016/j.nima.2006.07.066}{\emph{Nucl.\ Instrum.\
  Meth.\ A} {\bfseries 568} (2006) 839}.

\bibitem{ProcCalib2005}
{\scshape Pierre Auger} collaboration, \emph{{Calibration of the surface array
  of the Pierre Auger Observatory}},  in \emph{{29th International Cosmic Ray
  Conference}}, 8, 2005.

\bibitem{EPJ-C-MARTA}
P.~Abreu et~al., \emph{{MARTA: a high-energy cosmic-ray detector concept for
  high-accuracy muon measurement}},
  \href{https://doi.org/10.1140/epjc/s10052-018-5820-2}{\emph{Eur.\ Phys.\ J.\
  C} {\bfseries 78} (2018) 333}
  [\href{https://arxiv.org/abs/1712.07685}{{\ttfamily 1712.07685}}].

\bibitem{LL_RPC}
L.~Lopes et~al., \emph{{Outdoor Field Experience with Autonomous RPC Based
  Stations}},
  \href{https://doi.org/10.1088/1748-0221/11/09/C09011}{\emph{JINST} {\bfseries
  11} (2016) C09011} [\href{https://arxiv.org/abs/1606.03240}{{\ttfamily
  1606.03240}}].

\bibitem{JINST-PREC}
P.~Assis, P.~Brogueira, M.~Ferreira, R.~Luz and L.~Mendes, \emph{{Design and
  characterization of the PREC (Prototype Readout Electronics for Counting
  particles)}},
  \href{https://doi.org/10.1088/1748-0221/11/08/T08004}{\emph{JINST} {\bfseries
  11} (2016) T08004}.

\bibitem{CORSIKA}
D.~Heck, G.~Schatz, T.~Thouw, J.~Knapp and J.~N. Capdevielle, \emph{{CORSIKA: A
  Monte Carlo code to simulate extensive air showers}}, {\emph{FZKA-6019}
  (1998) }.

\bibitem{SimHernan}
H.~Asorey, L.~A. Núñez and M.~Suárez-Durán, \emph{Preliminary results from
  the latin american giant observatory space weather simulation chain},
  \href{https://doi.org/10.1002/2017SW001774}{\emph{Space Weather} {\bfseries
  16} (2018) 461}
  [\href{https://arxiv.org/abs/https://agupubs.onlinelibrary.wiley.com/doi/pdf/10.1002/2017SW001774}{{\ttfamily
  https://agupubs.onlinelibrary.wiley.com/doi/pdf/10.1002/2017SW001774}}].

\bibitem{Offline}
S.~Argir\'o, S.~L.~C. Barroso, J.~Gonzalez, L.~Nellen, T.~C. Paul, T.~A. Porter
  et~al., \emph{{The Offline Software Framework of the Pierre Auger
  Observatory}}, \href{https://doi.org/10.1016/j.nima.2007.07.010}{\emph{Nucl.\
  Instrum.\ Meth.\ A} {\bfseries 580} (2007) 1485}
  [\href{https://arxiv.org/abs/0707.1652}{{\ttfamily 0707.1652}}].

\bibitem{Geant4-1}
{\scshape \textsc{Geant4}} collaboration, \emph{Geant4: A simulation toolkit},
  {\emph{Nucl.\ Instrum.\ Meth.\ A} {\bfseries 506} (2003) 250}.

\bibitem{Geant4-2}
{\scshape \textsc{Geant4}} collaboration, \emph{Geant4 developments and
  applications}, {\emph{IEEE Trans.\ Nucl.\ Sci.} {\bfseries 53} (2006) 270}.

\bibitem{Allison:2016lfl}
{\scshape \textsc{Geant4}} collaboration, \emph{{Recent developments in
  \textsc{Geant4}}},
  \href{https://doi.org/10.1016/j.nima.2016.06.125}{\emph{Nucl.\ Instrum.\
  Meth.\ A} {\bfseries 835} (2016) 186}.

\bibitem{Creusot:2010xe}
A.~Creusot and D.~Veberi\v{c}, \emph{{Simulation of large photomultipliers for
  experiments in astroparticle physics}},
  \href{https://doi.org/10.1016/j.nima.2009.11.023}{\emph{Nucl.\ Instrum.\
  Meth.\ A} {\bfseries 613} (2010) 145}
  [\href{https://arxiv.org/abs/1001.1283}{{\ttfamily 1001.1283}}].

\bibitem{RPCmeas}
P.~Fonte, \emph{Analytical calculation of the charge spectrum generated by
  ionizing particles in resistive plate chambers at low gas gain},
  \href{https://doi.org/10.1088/1748-0221/8/04/p04017}{\emph{Journal of
  Instrumentation} {\bfseries 8} (2013) P04017}.

\bibitem{Water:1981}
R.~C. Smith and K.~S. Baker, \emph{{Optical properties of the clearest natural
  waters (200-800 nm)}}, {\emph{Applied Optics} {\bfseries 20} (1981) 177}.

\bibitem{Tyvek:1999}
A.~Filevich, P.~Bauleo, H.~Bianchi, J.~R. Martino and G.~Torlasco,
  \emph{{Spectral-directional reflectivity of Tyvek immersed in water}},
  {\emph{Nucl.\ Instrum.\ Meth.\ A} {\bfseries 423} (1999) 108}.

\end{thebibliography}\endgroup

\begin{center}
\rule{0.2\columnwidth}{0.5pt}
\raisebox{-0.4ex}{\scriptsize$\bullet$}
\rule{0.2\columnwidth}{0.5pt}
\end{center}

\section*{The Pierre Auger Collaboration}

\begin{sloppypar}
A.~Aab$^{75}$,
P.~Abreu$^{67}$,
M.~Aglietta$^{50,49}$,
J.M.~Albury$^{12}$,
I.~Allekotte$^{1}$,
A.~Almela$^{8,11}$,
J.~Alvarez Castillo$^{63}$,
J.~Alvarez-Mu\~niz$^{74}$,
R.~Alves Batista$^{75}$,
G.A.~Anastasi$^{58,49}$
% ...
\end{sloppypar}


\begin{description}[labelsep=0.2em,align=right,labelwidth=0.7em,labelindent=0em,leftmargin=2em,noitemsep]
\item[$^{1}$] Centro At\'omico Bariloche and Instituto Balseiro (CNEA-UNCuyo-CONICET), San Carlos de Bariloche, Argentina
\item[$^{2}$] Centro de Investigaciones en L\'aseres y Aplicaciones, CITEDEF and CONICET, Villa Martelli, Argentina
\item[$^{3}$] Departamento de F\'\i{}sica and Departamento de Ciencias de la Atm\'osfera y los Oc\'eanos, FCEyN, Universidad de Buenos Aires and CONICET, Buenos Aires, Argentina
\item[$^{4}$] IFLP, Universidad Nacional de La Plata and CONICET, La Plata, Argentina
\item[$^{5}$] Instituto de Astronom\'\i{}a y F\'\i{}sica del Espacio (IAFE, CONICET-UBA), Buenos Aires, Argentina
\item[$^{6}$] Instituto de F\'\i{}sica de Rosario (IFIR) -- CONICET/U.N.R.\ and Facultad de Ciencias Bioqu\'\i{}micas y Farmac\'euticas U.N.R., Rosario, Argentina
\item[$^{7}$] Instituto de Tecnolog\'\i{}as en Detecci\'on y Astropart\'\i{}culas (CNEA, CONICET, UNSAM), and Universidad Tecnol\'ogica Nacional -- Facultad Regional Mendoza (CONICET/CNEA), Mendoza, Argentina
\item[$^{8}$] Instituto de Tecnolog\'\i{}as en Detecci\'on y Astropart\'\i{}culas (CNEA, CONICET, UNSAM), Buenos Aires, Argentina
\item[$^{9}$] Observatorio Pierre Auger, Malarg\"ue, Argentina
\item[$^{10}$] Observatorio Pierre Auger and Comisi\'on Nacional de Energ\'\i{}a At\'omica, Malarg\"ue, Argentina
\item[$^{11}$] Universidad Tecnol\'ogica Nacional -- Facultad Regional Buenos Aires, Buenos Aires, Argentina
\item[$^{12}$] University of Adelaide, Adelaide, S.A., Australia
\item[$^{13}$] Universit\'e Libre de Bruxelles (ULB), Brussels, Belgium
\item[$^{14}$] Vrije Universiteit Brussels, Brussels, Belgium
\item[$^{15}$] Centro Brasileiro de Pesquisas Fisicas, Rio de Janeiro, RJ, Brazil
\item[$^{16}$] Centro Federal de Educa\c{c}\~ao Tecnol\'ogica Celso Suckow da Fonseca, Nova Friburgo, Brazil
\item[$^{17}$] Universidade de S\~ao Paulo, Escola de Engenharia de Lorena, Lorena, SP, Brazil
\item[$^{18}$] Universidade de S\~ao Paulo, Instituto de F\'\i{}sica de S\~ao Carlos, S\~ao Carlos, SP, Brazil
\item[$^{19}$] Universidade de S\~ao Paulo, Instituto de F\'\i{}sica, S\~ao Paulo, SP, Brazil
\item[$^{20}$] Universidade Estadual de Campinas, IFGW, Campinas, SP, Brazil
\item[$^{21}$] Universidade Estadual de Feira de Santana, Feira de Santana, Brazil
\item[$^{22}$] Universidade Federal do ABC, Santo Andr\'e, SP, Brazil
\item[$^{23}$] Universidade Federal do Paran\'a, Setor Palotina, Palotina, Brazil
\item[$^{24}$] Universidade Federal do Rio de Janeiro, Instituto de F\'\i{}sica, Rio de Janeiro, RJ, Brazil
\item[$^{25}$] Universidade Federal do Rio de Janeiro (UFRJ), Observat\'orio do Valongo, Rio de Janeiro, RJ, Brazil
\item[$^{26}$] Universidade Federal Fluminense, EEIMVR, Volta Redonda, RJ, Brazil
\item[$^{27}$] Universidad de Medell\'\i{}n, Medell\'\i{}n, Colombia
\item[$^{28}$] Universidad Industrial de Santander, Bucaramanga, Colombia
\item[$^{29}$] Charles University, Faculty of Mathematics and Physics, Institute of Particle and Nuclear Physics, Prague, Czech Republic
\item[$^{30}$] Institute of Physics of the Czech Academy of Sciences, Prague, Czech Republic
\item[$^{31}$] Palacky University, RCPTM, Olomouc, Czech Republic
\item[$^{32}$] Universit\'e Paris-Saclay, CNRS/IN2P3, IJCLab, Orsay, France, France
\item[$^{33}$] Laboratoire de Physique Nucl\'eaire et de Hautes Energies (LPNHE), Universit\'es Paris 6 et Paris 7, CNRS-IN2P3, Paris, France
\item[$^{34}$] Univ.\ Grenoble Alpes, CNRS, Grenoble Institute of Engineering Univ.\ Grenoble Alpes, LPSC-IN2P3, 38000 Grenoble, France, France
\item[$^{35}$] Bergische Universit\"at Wuppertal, Department of Physics, Wuppertal, Germany
\item[$^{36}$] Karlsruhe Institute of Technology, Institute for Experimental Particle Physics (ETP), Karlsruhe, Germany
\item[$^{37}$] Karlsruhe Institute of Technology, Institut f\"ur Kernphysik, Karlsruhe, Germany
\item[$^{38}$] Karlsruhe Institute of Technology, Institut f\"ur Prozessdatenverarbeitung und Elektronik, Karlsruhe, Germany
\item[$^{39}$] RWTH Aachen University, III.\ Physikalisches Institut A, Aachen, Germany
\item[$^{40}$] Universit\"at Hamburg, II.\ Institut f\"ur Theoretische Physik, Hamburg, Germany
\item[$^{41}$] Universit\"at Siegen, Fachbereich 7 Physik -- Experimentelle Teilchenphysik, Siegen, Germany
\item[$^{42}$] Gran Sasso Science Institute, L'Aquila, Italy
\item[$^{43}$] INFN Laboratori Nazionali del Gran Sasso, Assergi (L'Aquila), Italy
\item[$^{44}$] INFN, Sezione di Catania, Catania, Italy
\item[$^{45}$] INFN, Sezione di Lecce, Lecce, Italy
\item[$^{46}$] INFN, Sezione di Milano, Milano, Italy
\item[$^{47}$] INFN, Sezione di Napoli, Napoli, Italy
\item[$^{48}$] INFN, Sezione di Roma ``Tor Vergata'', Roma, Italy
\item[$^{49}$] INFN, Sezione di Torino, Torino, Italy
\item[$^{50}$] Osservatorio Astrofisico di Torino (INAF), Torino, Italy
\item[$^{51}$] Politecnico di Milano, Dipartimento di Scienze e Tecnologie Aerospaziali , Milano, Italy
\item[$^{52}$] Universit\`a del Salento, Dipartimento di Matematica e Fisica ``E.\ De Giorgi'', Lecce, Italy
\item[$^{53}$] Universit\`a dell'Aquila, Dipartimento di Scienze Fisiche e Chimiche, L'Aquila, Italy
\item[$^{54}$] Universit\`a di Catania, Dipartimento di Fisica e Astronomia, Catania, Italy
\item[$^{55}$] Universit\`a di Milano, Dipartimento di Fisica, Milano, Italy
\item[$^{56}$] Universit\`a di Napoli ``Federico II'', Dipartimento di Fisica ``Ettore Pancini'', Napoli, Italy
\item[$^{57}$] Universit\`a di Roma ``Tor Vergata'', Dipartimento di Fisica, Roma, Italy
\item[$^{58}$] Universit\`a Torino, Dipartimento di Fisica, Torino, Italy
\item[$^{59}$] Benem\'erita Universidad Aut\'onoma de Puebla, Puebla, M\'exico
\item[$^{60}$] Centro de Investigaci\'on y de Estudios Avanzados del IPN (CINVESTAV), M\'exico, D.F., M\'exico
\item[$^{61}$] Unidad Profesional Interdisciplinaria en Ingenier\'\i{}a y Tecnolog\'\i{}as Avanzadas del Instituto Polit\'ecnico Nacional (UPIITA-IPN), M\'exico, D.F., M\'exico
\item[$^{62}$] Universidad Aut\'onoma de Chiapas, Tuxtla Guti\'errez, Chiapas, M\'exico
\item[$^{63}$] Universidad Nacional Aut\'onoma de M\'exico, M\'exico, D.F., M\'exico
\item[$^{64}$] Institute of Nuclear Physics PAN, Krakow, Poland
\item[$^{65}$] University of \L{}\'od\'z, Faculty of Astrophysics, \L{}\'od\'z, Poland
\item[$^{66}$] University of \L{}\'od\'z, Faculty of High-Energy Astrophysics,\L{}\'od\'z, Poland
\item[$^{67}$] Laborat\'orio de Instrumenta\c{c}\~ao e F\'\i{}sica Experimental de Part\'\i{}culas -- LIP and Instituto Superior T\'ecnico -- IST, Universidade de Lisboa -- UL, Lisboa, Portugal
\item[$^{68}$] ``Horia Hulubei'' National Institute for Physics and Nuclear Engineering, Bucharest-Magurele, Romania
\item[$^{69}$] Institute of Space Science, Bucharest-Magurele, Romania
\item[$^{70}$] University Politehnica of Bucharest, Bucharest, Romania
\item[$^{71}$] Center for Astrophysics and Cosmology (CAC), University of Nova Gorica, Nova Gorica, Slovenia
\item[$^{72}$] Experimental Particle Physics Department, J.\ Stefan Institute, Ljubljana, Slovenia
\item[$^{73}$] Universidad de Granada and C.A.F.P.E., Granada, Spain
\item[$^{74}$] Instituto Galego de F\'\i{}sica de Altas Enerx\'\i{}as (IGFAE), Universidade de Santiago de Compostela, Santiago de Compostela, Spain
\item[$^{75}$] IMAPP, Radboud University Nijmegen, Nijmegen, The Netherlands
\item[$^{76}$] KVI -- Center for Advanced Radiation Technology, University of Groningen, Groningen, The Netherlands
\item[$^{77}$] Nationaal Instituut voor Kernfysica en Hoge Energie Fysica (NIKHEF), Science Park, Amsterdam, The Netherlands
\item[$^{78}$] Stichting Astronomisch Onderzoek in Nederland (ASTRON), Dwingeloo, The Netherlands
\item[$^{79}$] Universiteit van Amsterdam, Faculty of Science, Amsterdam, The Netherlands
\item[$^{80}$] Case Western Reserve University, Cleveland, OH, USA
\item[$^{81}$] Colorado School of Mines, Golden, CO, USA
\item[$^{82}$] Department of Physics and Astronomy, Lehman College, City University of New York, Bronx, NY, USA
\item[$^{83}$] Louisiana State University, Baton Rouge, LA, USA
\item[$^{84}$] Michigan Technological University, Houghton, MI, USA
\item[$^{85}$] New York University, New York, NY, USA
\item[$^{86}$] Pennsylvania State University, University Park, PA, USA
\item[$^{87}$] University of Chicago, Enrico Fermi Institute, Chicago, IL, USA
\item[$^{88}$] University of Delaware, Department of Physics and Astronomy, Bartol Research Institute, Newark, DE, USA
\item[] -----
\item[$^{a}$] School of Physics and Astronomy, University of Leeds, Leeds, United Kingdom
\item[$^{b}$] Max-Planck-Institut f\"ur Radioastronomie, Bonn, Germany
\item[$^{c}$] Fermi National Accelerator Laboratory, USA
\item[$^{d}$] also at Universidade Federal de Alfenas, Po\c{c}os de Caldas, Brazil
\item[$^{e}$] Colorado State University, Fort Collins, CO, USA
\item[$^{f}$] now at Hakubi Center for Advanced Research and Graduate School of Science, Kyoto University, Kyoto, Japan
\item[$^{g}$] also at Karlsruhe Institute of Technology, Karlsruhe, Germany
\item[$^{h}$] also at Radboud Universtiy Nijmegen, Nijmegen, The Netherlands
\end{description}

\end{document}